\documentclass{article}
\usepackage{amssymb}
\usepackage{amsmath,bm,float,graphicx}
\numberwithin{equation}{section}

\usepackage{graphicx}

\newtheorem{theorem}{Theorem}[section]

\newtheorem{remark}[theorem]{Remark}

\newcommand{\D}{\mathrm{d}}
\newcommand{\tr}{{\mathrm{tr}}}

\newcommand{\uu}{{\bf{u}}}
\newcommand{\ux}{{\bf{x}}}

\newcommand{\pp}[2]{ \frac{\partial #1}{\partial #2} }
\begin{document}
\title{On pressureless Euler equation with external force}
\author{B. G. Konopelchenko $^{1}$
and G.Ortenzi $^{2}$ 
\footnote{Corresponding author. E-mail: giovanni.ortenzi@unito.it }\\
$^1$ {\footnotesize INFN, Sezione di Lecce, via per Arnesano, 73100 Lecce, Italy} \\
$^2$ {\footnotesize  Dipartimento di Matematica ``G. Peano'', 
Universit\`{a} di Torino, via Carlo Alberto 10, 10123, Torino, Italy}\\
$^2${\footnotesize 
INFN, sezione di Torino,
via Pietro Giuria 1, 10125 Torino, Italy}
} 
\maketitle
\abstract{
Hodograph equations for the $n$-dimensional Euler equations with the constant pressure and external force linear in velocity are presented. 
They provide us with solutions of the Euler in implicit form and information on existence or absence of gradient catastrophes. 
It is shown that in even dimensions the constructed  solutions are periodic in time for particular subclasses of external forces.
Several particular examples in one, two and three dimensions are considered, including the case of Coriolis external force.
}
\section{Introduction}
The Euler and Navier-Stokes equations are the basic equations of hydrodynamics and theory of continuous media \cite{Lamb,L-VI}. These simple-looking
equations represent, in fact, a challenge as far as the analysis of their properties is concerned \cite{Lamb,L-VI,Whi}. 
Various approximations of the Euler and Navier Stokes equations and model equations resembling them have been discussed in a number of papers. \par

One of the approaches is to mimic the effects described by the r.h.s's of these equations by the external force depending on the velocity $\uu$, namely,
to  consider the equation (see e.g. \cite{Whi, Che91, KM96, CF03, Che19}) 
\begin{equation}
\rho \uu_t+ \rho \uu \cdot \nabla \uu = \mathbf{F}(t,\ux,\uu).
\label{HEeq-s}
\end{equation}
plus the continuity equation for the density $\rho$.\par

In the present paper we address the equation (\ref{HEeq-s}) 
the force $F_i=\rho (g_i+\sum_{k=1}^n A_{ik} u_k)$, i.e. the equation
\begin{equation}
 \pp{\uu_i}{t}+ \sum_{k=1}^n u_k \pp{u_i}{x_k} =  g_i+\sum_{k=1}^n A_{ik} u_k\, , \qquad i=1,\dots,n
\label{HEeq-lins-intro}
\end{equation}
where $g_i$ and $A_{ik}$ are constants. For $n=1$ and $A_{ik}=0$ this equation has been considered in \cite{CH,CF03}. The case 
$A_{ik}=A \delta_{ik}$  where $\delta_{ik}$ is the Kronecker-symbol has been discussed in \cite{Whi,Siv83,KM96,Kuz03,Che91,Che19}. \par

We demonstrate that the solutions of the equations (\ref{HEeq-lins-intro}) in the generic case $\det A  \neq 0$ are provided by the following hodograph equations
\begin{equation}
x_i + \sum_{m=0}^n \left( A^{-2} \left( e^{-t A}-1\right) \right)_{im}  \left( g_m+\sum_{l=1}^{n} A_{ml}u_l \right)
+ \sum_{m=0}^n  \left( A^{-1} \right)_{im} g_m t-\phi_i\left( \mathbf{M} \right)=0\, , \qquad i=1,\dots,n
\label{hodo-HElin-intro}
\end{equation}
and 
$\mathbf{M} $ has the following $n$ components
\begin{equation}
M_k= \sum_{k=0}^n \left(e^{-t A}\right)_{ik} u_k + \sum_{k=0}^n \left( A^{-1}(e^{-tA}-1) \right)_{ik} g_k  \, , \qquad k=1,\dots,n\, .
\label{inidatvar-intro}
\end{equation}
Solutions $u_i(\ux,t)$ are given implicitly by the formula
\begin{equation}
 u_i(t,\ux) = \sum_{k=0}^n \left( A^{-1} \left( e^{t A}-1\right) \right)_{ik}g_k + 
\sum_{k=0}^n  \left( e^{t A} \right)_{ik} u^0_k \left(  \mathbf{N} \right) \, ,  \qquad 
\mathrm{where} \, \qquad \uu^0=\uu(0,\ux)\, .
\label{impsol-intro}
\end{equation}
and 
\begin{equation}
N_i=x_i + \sum_{m=0}^n \left( A^{-2} \left( e^{-t A}-1\right) \right)_{im}   \left( g_m+\sum_{l=1}^{n} A_{ml}u_l \right)+ \sum_{m=0}^n  \left( A^{-1} \right)_{im} g_m t\, , \qquad i=1,\dots,n\, .
\end{equation}
It is shown that the blow-ups of derivatives of $\uu$ occur on the $n$-dimensional blow-up hypersurface $\Gamma$ given by the equation
\begin{equation}
\det \left( \left( A^{-1} \left( e^{t A}-1\right)  \right)_{im}  + \pp{\phi_i(\mathbf{M})}{M_m}\right) =0\, .
\label{catsur-intro}
\end{equation}
The equation (\ref{catsur-intro}) defining $\Gamma$ has a rather complicated form in terms of variables $t,\uu$. It assumes a much simpler form in the 
variables $t,\mathbf{M}$. In fact, these variables are convenient for the analysis of all properties of the equation (\ref{HEeq-lins-intro}).  \par

It is observed the structure and the properties of solutions of the equation (\ref{HEeq-lins-intro}) are quite different in different dimensions $n$.
In particular, it is demonstrated that they have a remarkabl novel feature in even dimensions. Namely, solutions of Euler equation (\ref{HEeq-lins-intro})
with $\mathbf{g}=0$ and certain $A$ are periodic in time $t$: $\uu(t+{T},\ux)=\uu(t,\ux)$. Correspondingly, the blow-up hypersurface $\Gamma$ is
composed by an infinite number of sheets.\par

We consider the properties of the equations (\ref{HEeq-lins-intro}) and (\ref{inidatvar-intro})  in one, two and three dimensions.  Concrete examples with 
particular initial data are discussed. \par

The situation with degenerated matrix $A$ is analysed too. The three dimensional cases with the Coriolis force 
$\mathbf{F}= \rho(-\mathbf{g}-\boldsymbol{\omega} \times \uu)$ is of special interest. \par


The paper is organized as follows.  Some basic known facts on the integral hypersurfaces associated with the partial differential equations are recalled in section \ref{sec-int-surf}. Hodograph equations (\ref{hodo-HElin-intro}), (\ref{impsol-intro})  and formula (\ref{catsur-intro}) are derived in section 
\ref{sec-linsource}. One dimensional case is considered in section \ref{sec-1D}. $N$-dimensional equation (\ref{HEeq-lins-intro}) with 
$A_{ik}=A \delta_{ik}$, $i,k=1,\dots n$ is discussed in section \ref{sec-nD-diag}. The case of the external   force linear in $\mathbf{u}$ 
with  $\mathbf{g}=0$ and solutions periodic in time are studied in section \ref{sec-periodic}.
Three examples of equation (\ref{HEeq-lins-intro}) in two-dimensions are considered in section \ref{sec-2dCoriolis}. The case of the external force with 
degenerate matrix $A$ ($\det A=0$) is treated in section \ref{sec-degenA}. Three-dimensional equation (\ref{HEeq-s}) with Coriolis force $\mathbf{F}$
is analysed in section \ref{sec-3DCor}. In the Appendix \ref{app-gen1D} the general one -dimensional case is considered in details.


\section{Integral hypersurface and characteristics}
\label{sec-int-surf}
 Here we will recall some general basic and well-known facts (see e.g. \cite{CH}) in a form convenient to a further use. \par
 
 Integral hypersurface  is one of the central objects associated with the quasilinear equations in dimension $n$. 
 It is a $(n+1)-$dimensional hypersurface in the 
 $(2n+1)-$dimensional space with coordinates $(t,\ux,\uu)$ defined by the equation
 \begin{equation}
 S_i(t,\ux,\uu)=0\, ,\qquad i=1, \dots,n
 \label{gen-eik}
 \end{equation}
 such that the resolution of the system (\ref{gen-eik}) with respect to $\uu$ provides us with the solution of the equation under consideration. \par
 
 Since the total derivatives of $S_i(t,\ux,\uu)$ on the level set (\ref{gen-eik})  vanish, the 
 functions $S_i(t,\ux,\uu)$ obey a certain  system of linear PDEs  which in the case  of the equation (\ref{HEeq-s}) is
 \begin{equation}
 \pp{S_i}{t}+ \sum_{k=1}^n u_k \pp{S_i}{x_k}+ \sum_{k=1}^n \frac{F_k}{\rho} \pp{S_i}{u_k}=0\, , \qquad i=1,\dots,n\, ,
 \label{eik-eq}
 \end{equation}
 where $(t,\ux,\uu)$ are independent variables. \par
 
 Indeed, let the functions $S_i(t,\ux,\uu)$ are such that a solution $\uu(t,\ux)$ of the equation (\ref{eik-eq}) is the solution of the equation (\ref{HEeq-s}).
 Since the total derivatives of $S_i(t,\ux,\uu(t,\ux))$ vanish one has
 \begin{equation}
 \begin{split}
 0=&\frac{\D S_i}{\D t}=\pp{S_i}{t}+\sum_{k=1}^n \pp{S_i}{u_k} \pp{u_k}{t}= 
 \pp{S_i}{t}+\sum_{k=1}^n \pp{S_i}{u_k} \left( - \sum_{l=1}^n u_l \pp{u_k}{x_l}+ \frac{F_k}{\rho}\right)\, , \qquad i=1,\dots,n \, ,\\
  0=&\frac{\D S_i}{\D x_k}=\pp{S_i}{x_k}+\sum_{l=1}^n \pp{S_i}{u_l} \pp{u_l}{x_k}\, ,   \qquad i,k=1,\dots,n\, .
 \end{split}
 \label{Dfront}
 \end{equation}
 Combining equations (\ref{Dfront}), one gets (\ref{eik-eq}). \par 

Any solution of the system (\ref{eik-eq}) provides us via (\ref{gen-eik}) with a local solution of the Euler equation (\ref{HEeq-s}) under the assumption that  that 
$\partial S_i/\partial u_j $ is invertible. 
Set of arbitrary functions $\phi_i(\mathbf{S}^{(1)},\dots \mathbf{S}^{(m)})$ of $m$ solutions $\mathbf{S}^{(i)}$ of (\ref{eik-eq}) is again a solution of the system (\ref{eik-eq}).
General solution of the system (\ref{eik-eq}) depends on $n$ arbitrary functions of $2n$ variables. 
The use of such general solution in (\ref{gen-eik}) gives a general solution of the of the Euler equation (\ref{HEeq-s}).\par

Equations (\ref{gen-eik}), allowing to find solutions of system of PDEs in an ``algebraic'' way, are, in fact, the hodograph equations in a general sense.\par

There are different methods to construct solutions of the system (\ref{eik-eq}), e.g. separation of variables, different Ansatz, etc.
One of the standard ways to do this  is to use the methods of characteristics. Characteristics for the system (\ref{eik-eq})
are defined by the equations
 \begin{equation}
 \frac{\D t}{\D \tau} =1\, , \qquad 
 \frac{\D x_i}{\D \tau} =u_i\, , \qquad 
 \frac{\D u_i}{\D \tau} =\frac{1}{\rho}F_i(t,\ux,\uu) \, , \qquad i=1,\dots,n \, .
\label{chargen}
 \end{equation}
Solutions $S_i$ of the system (\ref{eik-eq}) are constants along characteristics 
 \begin{equation}
 \frac{\D S_i}{\D \tau} =0 \, , \qquad i=1,\dots,n \, ,
\label{Rieminv-eq}
 \end{equation}
 i.e. they are integrals of motion for dynamical system (\ref{chargen}). If $I_1, \dots, I_m$ are independent 
 integrals of motion for the system (\ref{eik-eq}) then the functions 
 \begin{equation}
 S_i=\phi_i(I_1,\dots,I_m)\, , \qquad i=1,\dots,n\, ,
 \end{equation}
where $\phi_i$ are arbitrary functions, are solutions of the system (\ref{eik-eq}) which provide us with the solutions of the Euler equation via (\ref{gen-eik}). 
Different choices of the functions $\phi_i$ give different solutions of the equation (\ref{HEeq-s}). \par

In particular, choosing $S_i=I_k^{(i)}$, $I=1,\dots,n$ one gets particular solutions of Euler equations.

One constructs a general solution of the system (\ref{eik-eq})
if the dynamical system (\ref{chargen}) has $2n$ independent integrals of motion that is the system (\ref{chargen}) is maximally superintegrable.
(see e.g. \cite{MPW13,SUPERINT} and references therein). \par

In the case of the homogeneous Euler equation (HEE)  ($F_i \equiv 0$, $i=1,\dots,n$) there are $2n$ obvious integrals of motion
\begin{equation}
x_i-u_i t , \qquad u_i\, , \qquad i=1, \dots,n\, .
\end{equation}
So, the integral  hypersurface is given by the equation
\begin{equation}
\phi_i(\uu,\ux-\uu t)=0\, , \qquad  i=1, \dots,n\, ,
\label{HE-impsol}
\end{equation}
where $\phi_i$ are arbitrary functions of their arguments. \par

Resolving, when possible, the equation (\ref{HE-impsol}) with respect to the first $n$ arguments $\uu$, one obtains the standard formula
\begin{equation}
u_i=\phi_i(x_i-u_i t)\, ,  \qquad  i=1, \dots,n\, ,
\label{HE-charsol}
\end{equation}
for the solutions of HEE in implicit form, where $\phi_i$ are arbitrary functions. \par

Resolving instead, when possible, the equation (\ref{HE-impsol}) with respect to the second $n$ arguments $\ux-\uu t$, one gets 
\begin{equation}
x_i-u_i t=f_i( \uu)\, ,  \qquad  i=1, \dots,n\, ,
\label{HE-hodosol}
\end{equation}
where $f_i( \uu)$ are arbitrary functions, that is the standard form of hodograph equations for (HEE) \cite{KO22}.\par 

Some properties of the HEE, including the analysis of blowups of derivatives, have been studied in \cite{KO20para, KO22, KO23,KO24}.\par

 Method of characteristics, its generalizations and applications to partial differential equations have been discussed in several papers 
 ( see e. g. \cite{CH,Zel70, SZ89, Tsa91, Che91, Che19, Fai93, FL95, CF03, AF96, KM96, Kuz03, KM22, SZ07, ZS08, Zen11, KO20para, KO22, KO24}) .
In the present  paper we will derive the hodograph equations  for the system (\ref{HEeq-lins-intro})  and analyse the properties of their solutions. 
\section{Euler equation with external force linear in velocity}
\label{sec-linsource}
In the case of the external force $F_i=\rho \left( g_i+\sum_{k=1}^{n} A_{ik}u_k \right)$,  $i=1,\dots,n$ where $g_i$ and $A_{ik}$ are arbitrary constants,
 the Euler (\ref{HEeq-s}) equations assumes the form (\ref{HEeq-lins-intro})
\begin{equation}
\pp{u_i}{t} + \sum_{k=1}^n u_k \pp{u_i}{x_k}=g_i+\sum_{k=1}^{n} A_{ik}u_k\, ,  \qquad  i=1,\dots,n\, .
\label{HE-lins}
\end{equation}
The corresponding integral hypersurfaces  is defined by the solutions of the equation
\begin{equation}
\pp{S_i}{t} + \sum_{k=1}^n u_k \pp{S_i}{x_k}+\sum_{k=1}^{n} \Lambda_k \pp{S_i}{u_k}\, , \qquad \mathrm{with} \quad
\Lambda_k \equiv g_k+\sum_{l=1}^{n} A_{kl}u_l \, ,  \qquad  i=1,\dots,n\, ,
\label{eik-lins}
\end{equation}
and equations for charactestics are
\begin{equation}
\frac{\D t }{\D \tau}=1\, , \qquad 
\frac{\D x_i }{\D \tau}=u_i\, , \qquad 
\frac{\D u_i }{\D \tau}=\Lambda_i\, .
\label{charsys-linu}
\end{equation}
It is easy to check that the system (\ref{charsys-linu}) has $2n$ independent integrals of motion
\begin{eqnarray}
I^{(1)}_i&=&u_i- g_i t -\sum_{k=1}^{n} A_{ik}x_k\, , \qquad  i=1, \dots,n\,  \label{linvel-cq1}\, , \\
I^{(2)}_i&=&\sum_{k=1}^n\left(e^{-At}\right)_{ik} \Lambda_k  , \qquad  i=1, \dots,n\,  \label{linvel-cq2} \, .
\end{eqnarray}


 So, the integral hypersurface (\ref{gen-eik}) is defined by the equation 
 \begin{equation}
 \Phi_i(\mathbf{I^{(1)}},\mathbf{I^{(2)}})=0 \, , \qquad i=1,\dots,n\, ,
 \label{eik-lins-sol}
 \end{equation}
where $ \Phi_i$ are arbitrary functions and their arguments are given by the formulas (\ref{linvel-cq1}) and (\ref{linvel-cq2}). \par

Note that 
\begin{equation}
I^{(1)}_i (t=0)=u_i(x(0),0)- \sum_{k=1}^n A_{ik}  x_k(0) \, , \qquad 
I^{(2)}_i (t=0)=g_i+ \sum_{k=1}^n A_{ik} u_k(x(0),0) \, , \qquad 
i=1,\dots,n\, ,
\end{equation}
and  at $g_1 \to 0$ and $A \to 0$ 
\begin{equation}
I^{(1)}_i \Big{|}_ {g=0,A=0}=u_i \, , \qquad 
I^{(2)}_i \Big{|}_ {g=0,A=0}=0 \, , \qquad 
i=1,\dots,n\, ,
\end{equation}

For $\Phi_i$ resolvable with respect to the variable $u_i$, one obtains generic solutions of the equations (\ref{HE-lins}).\par

Considering functions $\mathbf{\tilde{I}^{(i)}}(\mathbf{I^{(1)}},\mathbf{I^{(2)}})$ for $i=1,2$ instead of $\mathbf{I^{(1)}},\mathbf{I^{(2)}}$,
one has other, but equivalent  forms of the hodograph equation (\ref{eik-lins-sol}).

Among all these possible choices the following  one ($\det A \neq 0$)
\begin{equation}
\begin{split}
M_i= &\sum_{k=1}^n \left( A^{-1} \right)_{ik} (I^{(2)}_k-g_k) 
=\sum_{k=1}^n \left( A^{-1} e^{-t A}\right)_{ik} \Lambda_k - \sum_{k=1}^n \left( A^{-1} \right)_{ik} g_k =\\
=&\sum_{k=1}^n \left(e^{-t A}\right)_{ik} u_k + \sum_{k=1}^n \left( A^{-1}(e^{-tA}-1) \right)_{ik} g_k \, ,
 \, , \qquad i=1, \dots, n
\end{split}
\label{cq-u0}
\end{equation}
and 
\begin{equation}
\begin{split}
N_i=& \sum_{k=1}^n  \left( \left( A^{-2} \right)_{ik} (I^{(2)}_k-g_k)  - \left( A^{-1} \right)_{ik} I^{(1)}_k  \right)=  \\
=&x_i+\sum_{k=1}^n \left( A^{-2} \left(e^{-t A}-1 \right)\right)_{ik} \Lambda_k  + \sum_{k=1}^n \left( A^{-1} \right)_{ik} g_k t \, , \qquad i=1, \dots, n
\end{split}
\label{lin-cq-xu}
\end{equation}
seems to be rather convenient when $\det(A )\neq 0$. Indeed, first one notes that 
\begin{equation}
\mathbf{M}(t=0)=\uu(\ux(0),0)\, , \qquad  \mathbf{N}(t=0)=\ux(0)\, .
\end{equation}
Further, in the limits $A \to 0$ and $g_i\to 0$ the formulae (\ref{lin-cq-xu})  reproduce the corresponding integrals for the equation (\ref{HE-lins})
with $A=0$ and $A=0,g_i=0$. Indeed, in the limit $A \to 0$
\begin{equation}
M_i=u_i-gt\, , \qquad N_i=x_i-u_i t + \frac{1}{2} g_it^2\,  , \qquad i=1, \dots, n\, ,
\end{equation}
i.e. the obvious integrals of motion of equation (\ref{chargen}) at $A=0$. For $g_i$ and $A=0$ one has the formulae (\ref{HE-impsol}),
 (\ref{HE-charsol}), and  (\ref{HE-hodosol}). \par
 
 So, the corresponding equation (\ref{gen-eik}) is of the form 
 \begin{equation}
 \psi_i(\mathbf{M},\mathbf{N})=0\, , \qquad i=1,\dots,n\, .
 \label{eik-lin-xu0}
 \end{equation}
where $\psi_i$ are arbitrary functions. Resolving it with respect to the variables $\mathbf{M}$, one has the implicit formula for $\uu$ 
\begin{equation}
\begin{split}
 u_i(t,\ux) = \sum_{k=0}^n \left( A^{-1} \left( e^{t A}-1\right) \right)_{ik}g_k + 
\sum_{k=0}^n  \left( e^{t A} \right)_{ik} u^0_k \left(  \mathbf{N} \right) \, ,  \qquad 
\mathrm{where} \, \qquad \uu^0=\uu(0,\ux)\, .
\end{split}
\label{implsol-HElin}
 \end{equation}
Resolution of (\ref{eik-lin-xu0}) with respect to $\mathbf{N}$, when possible, gives
\begin{equation}
x_i + \sum_{m=0}^n \left( A^{-2} \left( e^{-t A}-1\right) \right)_{im}  \Lambda_m
+ \sum_{m=1}^n  \left( A^{-1} \right)_{im} g_m t=\phi_i\left[ \left\{ \sum_{l=0}^n \left( A^{-1} e^{-t A} \right)_{kl} \Lambda_l -  \sum_{l=0}^n \left( A^{-1} \right)_{kl} g_l \right\}_{k=1,\dots,n}\right]
\label{hodo-HElin}
\end{equation}
where $\phi_i$, $i=1, \dots, n$ are arbitrary functions of $M_1, \dots, M_n$. Functions $\phi_i$ are connected with the initial data $\uu^0=\uu(t=0,\ux)$
for the equation (\ref{HE-lins}) by the relation $\ux=\boldsymbol{\phi}(\uu)$. It is the main reason for the choice of the hodograph 
equation in the form (\ref{eik-lin-xu0}) instead of (\ref{eik-lins-sol}). These hodograph equations 
look the most closed in the form to the equation (\ref{HE-hodosol}) for the HEE.\par

The formula (\ref{hodo-HElin}) provides us also with the information on the possible blow-ups of derivatives. Indeed, differentiating (\ref{hodo-HElin})
with respect to $x_l$ and $t$, one gets
\begin{equation}
\begin{split}
\delta_{il} & =\sum_{m=1}^n K_{im} \pp{u_m}{x_l}\, , \qquad i,l=1,\dots,n\,, \\
-u_i & = \sum_{m=1}^n K_{im} \left(  \pp{u_m}{t}- \Lambda_m \right)\, , \qquad i=1,\dots,n
\end{split}
\label{der-lin}
\end{equation}
where
 \begin{equation}
 K_{im} = \left( A^{-1} \left( 1- e^{-t A }\right) \right)_{im}+ \sum_{l=1}^n \pp{\phi_i}{u_l}  \left( e^{-t A }\right)_{lm}\, .
 \end{equation}
It is easy to see that if $\det K \neq 0$ then the relations (\ref{der-lin}) imply that $\uu(\ux,t)$ obey the equations (\ref{HE-lins}). \par

On the other hand if 
\begin{equation}
\det K =0
\label{bu-lincond}
\end{equation}
then the derivatives $\pp{u_i}{t}$ and $\pp{u_i}{x_k}$ blow up. Equation (\ref{bu-lincond}) defines the blow-up hypersurface $\Gamma$ in the 
hodograph space $(t,\uu)$. 

The formulae (\ref{cq-u0}) and (\ref{hodo-HElin}) indicate that the use of variables $\mathbf{M}$ instead of $\uu$ could simplify various formulae and calulations. Indeed, let us first express $\uu$ in terme of $\mathbf{M}$. The formula (\ref{cq-u0}) implies that  
\begin{equation}
u_i=\sum_{k=1}^n \left(e^{t A}\right)_{ik} M_k + \sum_{k=1}^n \left( A^{-1}(e^{tA}-1) \right)_{ik} g_k \, , \qquad i=1,\dots,n\, .
\end{equation}
Then
\begin{equation}
N_i=x_i+\sum_{k=1}^n \left( A^{-2} \left(1-e^{t A} \right)\right)_{ik} (g_k +\sum_{l=0}^n A_{kl}M_l)  +\sum_{k=1}^n \left( A^{-1} \right)_{ik} g_k t 
\, , \qquad i=1, \dots, n
\end{equation}
Consequently, instead of (\ref{implsol-HElin}) and (\ref{hodo-HElin}) one has 
\begin{equation}
M_i(t,\ux)=M_i^0(N_1, \dots, N_n)\, ,
\end{equation}
where $M_i^0(\ux)=u_i^0(\ux)=u_i(t=0,\ux)$.\\
The corresponding hodograph equation are
\begin{equation}
x_i+\sum_{k=0}^n \left( A^{-2} \left(1-e^{t A} \right)\right)_{ik} (g_k +\sum_{l=0}^n A_{kl}M_l)  +\sum_{k=0}^n \left( A^{-1} \right)_{ik} g_k t=
 \phi_i(M_1, \dots, M_n) \, , \qquad i=1,\dots,n,
\end{equation}
where $\boldsymbol{\phi}$ is the function inverse to $\mathbf{M}^0=\uu(t=0,\ux)$.\par

The functions $M_i$ obey the equations
\begin{equation}
\pp{M_i}{t}+ \sum_{k,l=0}^n \left( \left(e^{tA}\right)_{kl} M_l + \left( A^{-1} \left(e^{t A}-1 \right)\right)_{kl} g_l \right) \pp{M_i}{x_k}=0\, , \qquad i=1,\dots,n.
\end{equation}
 
The derivatives of $\uu$ and $\mathbf{M}$ blow up simultaneously since
\begin{equation}
\begin{split}
\pp{u_i}{x_k}=\sum_{l=0}^n \left(e^{tA}\right)_{il} \pp{M_l}{x_k}\, , \qquad   i=1,\dots,n\, , \\
\pp{u_i}{t}=\sum_{l=0}^n \left(Ae^{tA}\right)_{il} M_l+\sum_{l=0}^n \left(e^{tA}\right)_{il} \pp{M_l}{t}\, , \qquad   i=1,\dots,n\, .
\end{split}
\label{bu-uM}
\end{equation}
The blow-up hypersurface for $\mathbf{M}$ is given by a simple equation 
\begin{equation}
\det \left( \left( A^{-1} \left( e^{t A}-1\right)  \right)_{im}  + \pp{\phi_i(\mathbf{M})}{M_m}\right) =0\, .
\end{equation}

Note that for the homogeneous Euler equation  $M_i(t,\ux)=u_i(t,\ux)$, $i=1,\dots,n$.\par

Hodograph equations (\ref{hodo-HElin}) are convenient since they provide us with the class of solutions parametrized by the functions $\phi_i$
directly connected with the initial data $\uu(t,\ux)$. However there are solutions of the equation (\ref{HE-lins}) for calculations of which other forms of the 
 hodograph equations are more convenient. For example, the simple solution
 \begin{equation}
 u_i(t,\ux)=g_i t + \sum_{k=1}^n A_{ik} x_k\, , \qquad i=1,\dots,n 
 \end{equation}
of the equation (\ref{HE-lins}) is easily obtainable from the hodograph equation (\ref{eik-lins-sol}) with the functions $\boldsymbol{\Phi}$ not depending on $\mathbf{I}^{(2)}$. \par

Finally, we note that the degenerations of the functions $\boldsymbol{\Phi}$ in  (\ref{eik-lins-sol})  or $\boldsymbol{\psi}$ in  (\ref{eik-lin-xu0}) 
to those depending
only on one $n$-tuplet of the integrals of motions provides us with the simple solutions of the equation (\ref{HE-lins}). For instance, the hodograph
equation $\phi_i(\mathbf{I^{(1)}})=0$, $i=1,\dots,n$  implies that $I^{(1)}_i=\alpha_1$, $i=1,\dots,n$ where $\alpha_i$ are constants. Correspondingly
  one has the solution 
  \begin{equation}
  u_i(t,\ux)=g_i t + \sum_{k=1}^n A_{ik} x_k +\alpha_i\, , \qquad i=1,\dots,n\, .
  \end{equation}
Analogously, in the case when $I^{(2)}_i=\beta_i$ where $\beta_i$ are constants, one has
\begin{equation}
u_i(t,\ux)= \sum_{k=1}^n \left( A^{-1} e^{t A } \right)_{ik} \beta_k - \sum_{k=1}^n \left( A^{-1}\right)_{ik}g_k \, , \qquad i=1, \dots, n\, .
\label{sol-cq2-kost}
\end{equation}
When $M_i=\gamma_i$ where $\gamma_i$ are constants, the related solutions are
\begin{equation}
u_i=   \sum_{k=1}^n \left(  e^{tA} \right)_{ik} \gamma_k   + \sum_{k=1}^n \left( A^{-1}\left( e^{tA}-1 \right)\right)_{ik} g_k    \, , \qquad i=1, \dots, n \, .
\label{sol-cq1-kost}
\end{equation}
Finally, when $N_i=\delta_i$ where $\delta_i$ are constants,
\begin{equation}
u_i=  \sum_{k=1}^n \left[
\left( \left( e^{-tA}-1\right)^{-1}\right)_{ik} \left( \delta_k-x_k\right) + \left( \left( e^{-tA}-1\right)^{-1}\right)_{ik} g_k t - \left( A^{-1}\right)_{ik}g_k \right] \, , \qquad i=1, \dots, n \, .
\end{equation}
\section{One dimensional Euler equation}
\label{sec-1D}
In the one-dimensional case, the hodograph equation (\ref{hodo-HElin}) when $A \neq 0$ is 
\begin{equation}
x+A^{-1}gt+A^{-2} (e^{-tA}-1)(g+Au)=   \phi \left( A^{-1 }e^{-tA} (g+Au) - A^{-1}g \right)\, .
\label{hodo-lin1D}
\end{equation}
One also has
\begin{equation}
\begin{split}
\pp{u}{x} = &\frac{1}{ A^{-1}(1-e^{-tA})+ e^{-tA} \phi'\Big( A^{-1 }e^{-tA} (g+Au) - A^{-1}g \Big) } \, ,\\
\pp{u}{t} = &g+Au-\frac{u}{ A^{-1}(1-e^{-tA})+ e^{-tA} \phi'\Big( A^{-1 }e^{-tA} (g+Au) - A^{-1}g \Big)}\, , 
\end{split}
\end{equation}
and the blowup curve is defined by the equation
\begin{equation}
A^{-1}(e^{tA}-1)+ \phi'\left( A^{-1 }e^{-tA} (g+Au) - A^{-1}g \right)=0\, .
\end{equation}
In the limit $A,g \to 0$ one has the classical text-books example $t+\phi'(u)=0$ (see e.g. \cite{Whi}).  \par

In the case $g\neq 0$ and $A\to 0$ one has one dimensional gravitational force $F=\rho g$. Hodograp equation
becomes
\begin{equation}
x-ut +\frac{1}{2}gt^2 -   \phi \left( u-gt \right)=0\, .
\label{hodo-lin1D-g}
\end{equation}
One has
\begin{equation}
u  =  gt + u_0 \left( x-ut+ \frac{1}{2}gt^2\right)\, .
\label{sol-lin1D-g}
\end{equation}
The blow-up curve is defined by the equation 
\begin{equation}
t+ \phi' \left( u-gt \right)=0
\label{BU-1D-g}
\end{equation}
For $g=1$ the formula (\ref{hodo-lin1D-g}) is equivalent to that given long time ago in \cite{CH} Ch. II formula (9). The case case $g \neq 1$ is trivially connected to that with $g=1$ by the transformation $u\to u/g$ and $x \to x/g$. \par

Let us consider the blowups of the solutions of equation (\ref{HE-lins}). The catastrophe (positive minimum of the blow-up
curve (\ref{BU-1D-g}))  is characterized, generically, by the equations
\begin{equation}
\phi''(u_c-gt_c)=0\, , \qquad t_c+\phi'(u_c-gt_c)=0\, , \qquad x_c=u_ct _c-\frac{1}{2}gt_c^2 +   \phi \left( u_c-gt_c \right)\, , \qquad  \phi'''(u_c-gt_c)>0\, .
\end{equation}
The first relation can be obtained by the relation $t_u \equiv  \frac{\phi''(u-g t)}{1-g \phi''(u-g t)} =0$.
Let us compare this system with the analogue system for the $g=0$ generic case
\begin{equation}
\phi''(u_c|_{g=0})=0\, , \qquad t_c|_{g=0}+\phi'(u_c|_{g=0})=0\, , \qquad x_c|_{g=0}=u_c|_{g=0}t _c|_{g=0} +   \phi \left( u_c|_{g=0} \right)\, , 
\qquad  \phi'''(u_c|_{g=0})>0\, .
\end{equation}
A direct inspection shows that 
\begin{equation}
t_c=t_c|_{g=0}\, , \qquad u_c=u_c|_{g=0} + g t_c |_{g=0}\, , \qquad x_c=x_c|_{g=0}+ \frac{1}{2}g(t_c|_{g=0})^2\, .
\end{equation}
We remark that,  until the blowup, in the homogeneous case the extremal values of $u$ does not evolve in time while the presence of a source allows 
a variation on time of such values.\par

For the standard text-book initial data
\begin{equation}
u(t=0,x)=\mu \left( 1-\tanh \left( \kappa x\right)\right)\, , \qquad \mu \in \mathbb{R},\quad \kappa \in \mathbb{R}_+\, ,
\label{condinitanh1D-g}
\end{equation}
the local inverse of the initial data is $\phi(u)=\frac{1}{\kappa}\, \mathrm{arctanh}\left(1-\frac{u}{\mu}\right)$ and the blow-up curve is given by the equation
\begin{equation}
t=\frac{1}{\kappa \mu} \frac{1}{1-\left(1- \frac{u- g t}{\mu}\right)^2}\, .
\end{equation}

In terms of the variables $t$ and $u$ it is the cubic curve
\begin{equation}
g^2 t^3 + 2g (\mu-u) t^2+(u^2-2 \mu u)t+\frac{\mu}{\kappa}=0\, .
\end{equation}

The equation for this blowup curve is much simpler in the variables $t$ and $M=u-gt$. Indeed, it is 
\begin{equation}
t=\frac{1}{\kappa \mu} \frac{1}{1-\left(1- \frac{M}{\mu}\right)^2}\, ,
\end{equation}
where $0 \leq M \leq 2 \mu$. Consequently the derivative of $M$ and $u$ blowup on this timple curve with $t_{\min}=\frac{1}{\kappa \mu}$ and 
the corresponding $u_{\min}=\mu+\frac{g}{\kappa \mu}$. Note that $t_{\min}$ coincides with that for $g=0$ while the corresponding $u_{\min}$ is shifted 
by $\frac{g}{\kappa \mu}$. One also has $ x_{\min}=\frac{1}{\kappa}+\frac{g }{2\kappa^2\mu^2}+\frac{1}{\kappa}\, \mathrm{arctanh}\left(-\frac{g }{\kappa \mu^2}\right)$.

\begin{figure}[h]
\begin{center}
\includegraphics[width=.4 \textwidth]{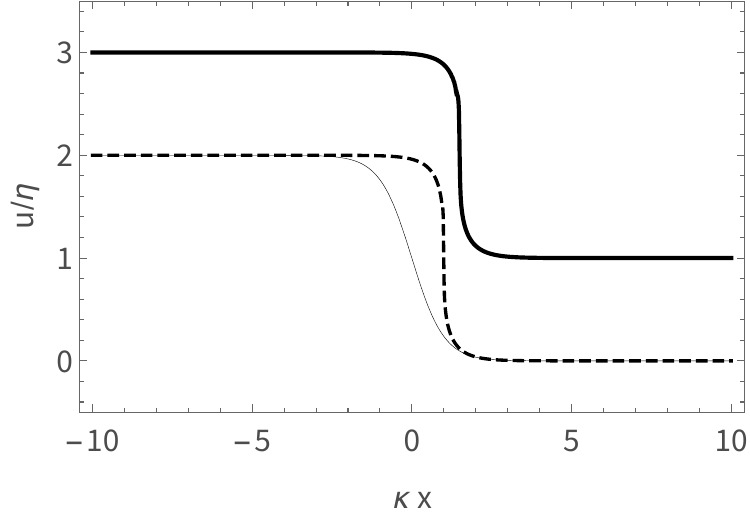}
\caption{Evolution of the initial datum $u=1-\tanh(x)$, corresponding to (\ref{condinitanh1D-g}) with $\mu=\kappa=1$. 
The thin line is the initial datum $u(x,t=0)$, the dashed line is $u(x,t_c=1)$ with $g=0$ while the bold line is $u(x,t_c=1)$ with $g=1$.}
\label{fig-exe-gravity-1D}
\end{center}
\end{figure}

In Figure \ref{fig-exe-gravity-1D} is shown the different evolution of the cases $g=0$ and $g\neq0$. 
 
In the general case $g \neq 0$ and $A \neq 0$  the situation is more intriguing. Solutions of the Euler equation implicitly are given by the formula
 \begin{equation}
 u(t,x)=A^{-1} \left(e^{tA}-1\right)g+e^{tA} u_0 \left(x+ A^{-2} \left(e^{-tA}-1\right) \left(g+Au\right)+A^{-1} g t \right)\, .
 \end{equation}
The equation defining the blow-up is rather complicated even for simple initial data, but it is simplified drastically if one used the variables $t$
and $M$ (see (\ref{cq-u0})), i.e.
\begin{equation}
M=e^{-tA}u+A^{-1} \left(e^{-tA}-1\right) g\, .
\end{equation}
Note that $M_x=e^{-tA}u_x$ and therefore $M$ blows-up if and only if $u$ does.
The blowup condition assumes the form
\begin{equation}
t=\frac{1}{A}\log\left(1- A\pp{\phi(M)}{M} \right)\, .
\label{bu-cond-lin-1D}
\end{equation}
Note that the interval of variation of $M$ coincides with that of $u_0$ (see (\ref{bu-uM})).  
It is obvious that the equation (\ref{bu-cond-lin-1D}) has no real solutions for $t$ if
\begin{equation}
A \pp{\phi(M)}{M}>1\, ,
\label{nobu-cond-1D}
\end{equation}
 for the whole interval of variation of $M$. In this case the corresponding solutions of the one-dimensional Euler equation (\ref{HE-lins}) do not
 exhibit blowups. \par

Blowups occur if
\begin{equation}
A \pp{\phi(M)}{M}<1\, .
\end{equation}
Moreover if 
\begin{equation}
A \pp{\phi(M)}{M}<0\, .
\end{equation}
then the sign of the blow-ups time $t$ coincides with the sign of $A$. If, instead, 
\begin{equation}
0<A \pp{\phi(M)}{M}<1\, ,
\end{equation}
the sign of the time $t$ and $A$ are opposite each other.\par

It is of interest to compare the behavior of the solutions of the Euler equation with different values of $A$, but with the same initial data $u_0(x)$ 
(recall that $\phi(u_0(x))=x$). One immediately concludes that the existence or absence of blowups depends on the value of $A$. In the limit $A\to 0 $
one has
\begin{equation}
t_0 \equiv t|_{A=0}=-\pp{\phi}{u} 
\end{equation}
that gives the blowup time for one-dimensional Euler equation only with the gravitational force $\rho g$ in the r.h.s.\, .
Thus the condition (\ref{bu-cond-lin-1D}) can be written as
\begin{equation}
A t = \log \left( 1+ A t_0\right)\, .
\end{equation}

So the condition (\ref{nobu-cond-1D}) of the blowup absence for the solutions of the one dimensional equation (\ref{HE-lins}) with $A \neq 0$, but with 
the same $\phi(u)$ as in can be rewritten as
\begin{equation}
A\,  t_0 <-1.
\end{equation}
Hence, if $ t_0>0$ then the corresponding solution (with the same initial data) of the Euler equation with 
\begin{equation}
A<0, \qquad |A|>\frac{1}{t_0} 
\label{nobu-1D-A}
\end{equation}
is free of blowups. 
At $t_0<0$ the condition of absence of blowups is 
\begin{equation}
A>0, \qquad A>\frac{1}{\big{|}t_0\big{|}}\, . 
\end{equation}
Note that in the case $A\, t_0 >0$ the effect of the external force is only the change of the blowup time.
Some particular examples are considered in Appendix 
\ref{app-gen1D}. \par

It is noted that the hodograph equation and blowups condition for the one-dimensional equation (\ref{HE-lins}) with $g=0$ and $A\to -A$ have been
considered in \cite{Whi} \S 2.12 using the Lagrangian coordinates $\xi$ instead of the Eulerian one used here. They are related by the obvious identification 
$\xi=\phi(M)$. In the variable $\xi$ the condition (\ref{nobu-cond-1D}) becomes ($f(\phi(M))=1$)
\begin{equation}
\pp{f(M)}{\xi}<A
\end{equation}
that is the condition (2.77) in \cite{Whi} (with $g \neq 0$).\par

It should be noted also that the condition (\ref{nobu-1D-A}) of the absence of the blow-ups for the Euler-equation with $g=0$, $A<0$ and $t>0$
has been derived earlier in a different manner in \cite{Che19}. \par

Finally, we note that in terms of the variable $M$, the hodograph equation (\ref{hodo-lin1D}) is given by
\begin{equation}
x+ A^{-2} (1-e^{tA})g + A^{-1}gt + A^{-2} (1-e^{tA})M=\phi(M)
\label{hodo1D-slin}
\end{equation}
where 
\begin{equation}
M=e^{-tA}u+A^{-1}(e^{-tA}-a)g\, .
\end{equation}
Introducing the obvious new variables
\begin{equation}
\begin{split}
\overline{t}=&A^{-1} \left(e^{tA}-1\right)\, , \\
\overline{x}=&x+gA^{-2}\left(1-e^{tA}\right)+A^{-1}gt\, , \\
\overline{u}(\overline{t},\overline{x})=&M(t,x)=e^{-tA}u+gA^{-1}(e^{-tA}-1)\, ,
\end{split}
\label{snsmap}
\end{equation}
one rewrites the hodograph equation (\ref{hodo1D-slin}) as
\begin{equation}
\overline{x}-\overline{u}\, \overline{y}=\phi(\overline{u}) \, ,
\end{equation}
thet is, the classical hodograph equation for the one-dimensional homogeneous Euler equation (Burgers-Hopf equation) 
$\overline{u}_{\overline{t}}+\overline{u} \overline{u}_{\overline{x}}=0$. \par

In the case $g=0$, the transformation (\ref{snsmap}) has been discussed in  \cite{Che91,Che19}. \par

The defect of this transformation is that it is the mapping between the full interval $(-\infty,+\infty)$ for the variable
$t$ and semi-infinite intervals for $\overline{t}$, namely, the interval $[-A^{-1}, +\infty)$ for $A>0$  and the interval 
$[-\infty,|A|^{-1})$ for $A<0$.
\section{Multi-dimensional  Euler equation with external force $F=\rho (\mathbf{g}+A \uu)$}
\label{sec-nD-diag}
In this section we will consider the Euler equation (\ref{HE-lins}) with $A_{ik}=A \delta_{ik}$, $i,k=1,\dots,n$ where A is a constant.\par

Hodograph equation in this case is of the form
\begin{equation}
x_i + A^{-2} (e^{-tA}-1)(g_i+A u_i)+ A^{-1}g_i t-\phi_i \left( e^{-tA} \uu+ A^{-1} (e^{-tA}-1) \mathbf{g} \right)=0\, .
\label{hodo-nD-diag}
\end{equation}
The blow-up hypersurface is defined by the equation
\begin{equation}
\det \left( A^{-1} (e^{tA}-1)\delta_{ik} + \pp{\phi_i}{M_k}\right)=0
\end{equation}
where $\mathbf{M}=e^{-tA} \uu+ A^{-1} (e^{-tA}-1) \mathbf{g}$.
In the variable 
\begin{equation}
\tau \equiv A^{-1} (e^{tA}-1) 
\label{teff-nD-diag}
\end{equation}
this equation assumes the form
\begin{equation}
\det \left( \tau\delta_{ik} + \pp{\phi_i}{M_k}\right)=0
\label{bu-curve-nD-diag}
\end{equation}
which is exactly of the same form as for the HEE with the substitution $t \to \tau$ \cite{KO22}. \par

Equation (\ref{bu-curve-nD-diag}) is a polynomial equation of order $n$ in the variable $\tau$. If it has $m$ real roots $\tau_k=f_k(\mathbf{M})$, 
$k=1, \dots,m$, then the blowup hypersurface  $\Gamma_\tau$ is a union (similar to the HEE case \cite{KO22})
\begin{equation}
\Gamma_\tau=\cup_{k=1}^m \{ \tau_k= f_k(\mathbf{M})\}\, .
\end{equation}
Due to the relation (\ref{teff-nD-diag}) the blowup hypersurface $\Gamma$ in terms of the variable $t$ and $\uu$ is different from $\Gamma_\tau=0$.
Indeed, for each branch described above
\begin{equation}
e^{t_k A}=1+A f_k(\mathbf{M})\, \qquad k=1,\dots,k\, .
\end{equation}
So, $t_k$ are real-valued only if
\begin{equation}
A f_k(\mathbf{M})>-1\, ,
\label{reg-cond-diag}
\end{equation}
for all admissible values of $\mathbf{M}$. Consequently, the blowup hypersurface $\Gamma$ is a union 
\begin{equation}
\Gamma_\tau=\cup_{k=1}^{m^*} \{ t_k= A^{-1}\log(1+Af_k(\mathbf{M}))\}\, ,
\label{hsup-bu-tM-diag}
\end{equation}
where $m^*<m$ is a number of cases for which the condition (\ref{reg-cond-diag}) is satisfied. At $A=0$ we have $m^*=m$.\par

For given initial data (functions $\phi_k$) the number $m^*$ of branches of $\Gamma$ depends on the value of $A$. Typically, some $m_+$ of the
function $f_k(\mathbf{M})$ are positive for all values of $\mathbf{M}$ and others $m_-$ are negative. We denote the minimal value
of all positive functions $f_k$ as $f_{\min}$ and the maximal value of all negative functions $f_k$ as $f_{\max}$.\par

It is easy to see that, for positive  $f_k(\mathbf{M})$,  the condition (\ref{hsup-bu-tM-diag}) is satisfied for all positive $A$ and for negative $A$ obeying to the
 condition $|A| f_k(\mathbf{M})<1$ for all values of $\mathbf{M}$. For negative $f_k(\mathbf{M})$, the condition  (\ref{hsup-bu-tM-diag}) is satisfied for all 
 negative $A$ obeying the constraint  $A |f_k(\mathbf{M})|<1$. \par
 
 On the other hand if the constant $A$ is such that for some $k$ 
 \begin{equation}
 A f_k(\mathbf{M})<-1
 \label{nobu-diag}
 \end{equation}
then in (\ref{hsup-bu-tM-diag}) there is no contribution corresponding to these values of $k$.  The condition (\ref{nobu-diag}) is realizable for $A$ and 
$f_k(\mathbf{M})$ of opposite signs. For positive $f_k$ and negative $A$ the condition (\ref{nobu-diag})  is
\begin{equation}
|A|> \frac{1}{f_{\min}}
\end{equation}
while for negative $f_k$ and positive $A$ the condition (\ref{nobu-diag})  is
\begin{equation}
A> \frac{1}{|f_{\min}|}\, .
\end{equation}
thus, the presence of the external force $\mathbf{F}=\rho (\mathbf{g}+A \uu)$ with sufficiently large $A$ in modulo suppress the blowups either 
at positive time ($A<0$) or at negative time ($A>0$).\par

Generically, it does not regularize a solution of the corresponding pressureless Euler  equation at all values of time $t$. Complete regularization
is achieved in the particular cases when all functions $f_k(\mathbf{M})$, $k=1,\dots,m$ are positive or negative.  \par
 
For the illustration of the observations presented above we consider the two dimensional case with initial data 
(compare with the example (6.2) in \cite{KO22})
\begin{equation}
{u_1}_0=-\tanh(x_1+\epsilon x_2)\, , \qquad {u_2}_0=-\tanh(\epsilon x_1+ x_2) \, , \qquad \epsilon>0. 
\label{2D-exe-eps-diag}
\end{equation}
The corresponding functions $\phi_1$ and $\phi_2$ are
 \begin{equation}
 \phi_1(\mathbf{M})=\frac{1}{\epsilon^2-1} \left( \mathrm{atanh}(M_1) - \epsilon \,  \mathrm{atanh}(M_2) \right)\, , \qquad
  \phi_2(\mathbf{M})=\frac{1}{\epsilon^2-1} \left( - \epsilon \, \mathrm{atanh}(M_1) +  \mathrm{atanh}(M_2) \right)\, .
 \end{equation}
The equation (\ref{bu-curve-nD-diag}) is of the form
\begin{equation}
\tau ^2 +\frac{2-M_1^2-M_2^2}{(\epsilon^2-1)(1-M_1^2)(1-M_2^2)} \tau-\frac{1}{(\epsilon^2-1)(1-M_1^2)(1-M_2^2)}  =0\, ,
\label{bu-curve-2D-diag-tanh}
\end{equation}
where $0 \leq M_1,M_2 \leq 1$.  Equation (\ref{bu-curve-2D-diag-tanh}) has two real roots  $\tau_\pm$ for all $\epsilon>0$
\begin{equation}
\tau_\pm \equiv f\pm(\mathbf{M})=\frac{(2-M_1^2-M_2^2)\pm \sqrt{(2-M_1^2-M_2^2)^2+4(\epsilon^2-1)(1-M_1^2)(1-M_2^2)}}{2(\epsilon^2-1)(1-M_1^2)(1-M_2^2)}\, .
\end{equation}
It is easy to check that $f_+>0$ for $\epsilon >0$ while $f_->0$ for $0<\epsilon \leq 1$ and  $f_-<0$ for $\epsilon > 1$. So, for  $0<\epsilon \leq 1$
both $f_\pm(\mathbf{M})$ are positive while for $\epsilon >1$ they have opposite sign. 
In addition, at    $0<\epsilon \leq 1$ we have $f_+(\mathbf{M})< f_-(\mathbf{M})$ and  minimal value of $f_+$ is ${f_+}_{\min}=\frac{1}{1+\epsilon}$.
Maximal value of  $f_-$ is ${f_-}_{\max}=\frac{1}{1-\epsilon}$.\par 

Thus for $A$ obeying the condition (\ref{reg-cond-diag}) the blowup surface has two branches. If, instead, $A$ satisfies the opposite condition 
(\ref{nobu-diag}) the behavior of the solution depends on the value of $\epsilon$. At $\epsilon >1$ and negative $A$ such that
  \begin{equation}
  |A|>1+\epsilon
  \end{equation}
this solution as no blowup at $t>0$ (gradient catastrophe) and has blowup at negative time. For $A>0$ and $\epsilon>1$ such that
\begin{equation}
A>\epsilon-1
\end{equation}
the solution has no blowup at positive time.\par

Solution of the Euler equation with initial data (\ref{2D-exe-eps-diag}) and value of parameter $\epsilon$ in the interval $0\leq \epsilon \leq 1$ has quite
different properties. If the external force is such that $A<0$ and 
\begin{equation}
|A|>1+\epsilon
\end{equation}
this solution
 is free of blowups for allvalues of time $t$. \par
 Other examples of solutions of the 2D pressureless Euler equation have similar local properties.  \par
 
At last, we note that in terms of the variables
  \begin{equation}
 \overline{t} = A^{-1} \left( e^{tA}-1 \right)\, , \qquad
 \overline{\ux} = \ux -A^{-2} \left( e^{tA}-1 \right) \mathbf{g}+ A^{-1}  \mathbf{g} t\, , \qquad
 \overline{\uu} = \mathbf{M}= e^{-tA}\uu-A^{-1} \left(1- e^{-tA} \right)\mathbf{g}\, ,
 \end{equation}
 the hodograph equation become
 \begin{equation}
 \overline{\ux}-\overline{\uu}\overline{t}=\boldsymbol{\phi}(\overline{\uu})\, ,
 \end{equation}
and the function $\overline{\uu}(\overline{t},\overline{\ux})$ obeys the $n-$dimensional homogeneous  Euler equation.
For $\mathbf{g}=\mathbf{0}$ see also \cite{Che91,Che19}.
\section{Solutions periodic in time}
\label{sec-periodic}
The results for the particular external force presented in the previous section are quite direct extension of those in the one-dimensional case. \par

In this section we will demonstrate that solutions of the multidimensional equation (\ref{HEeq-lins-intro}) have a remarkable novel property which does not
exist in one dimensional case. Namely, we will show that in the even dimensions for external force ${F_i}=\rho \sum_{j=1}^n A_{ij} u_j$ 
(i.e. $\mathbf{g}=0$ case)  with a particular matrix $A$ the solutions of the equation (\ref{HEeq-lins-intro}) constructed via the hodograph equations
(\ref{hodo-HElin-intro})-(\ref{inidatvar-intro})  are periodic in time $\uu(t+{T},\ux)=\uu(t,\ux)$ with certain period ${T}$.\par

First, we observe that $\mathbf{g}=0$ the $t$-dependence of the solutions (\ref{implsol-HElin}) is determined by the structure of the matrix $e^{tA}$. In 
particular, it is easy to see that the periodicity of this matrix, i.e. $e^{(t+{T})A}=e^{tA}$ with some real ${T}$ is a necessary condition for the periodicity
of solutions $\uu(t,\ux)$.\par

So, solutions of the equation (\ref{HEeq-lins-intro}) constructed by the hodograph equations (\ref{hodo-HElin-intro}) will be nontrivially periodic in time only if the matrix $A$ is nondegenerated ($\det A \neq 0$ ) and there exists a real ${T}$ such that 
\begin{equation}
e^{{T} A}=1\, .
\label{percons}
\end{equation}
We restrict ourselves to the case of diagonalizable matrix $A$ with $B A B^{-1}={{A_d}}$ where $({A_d})_{ij}=\delta_{ij} \nu_j$, $i,j=1,\dots,n$ and 
$\nu_j$ are eigenvalues of $A$. In this case the constraint (\ref{percons}) is equivalent to the following
\begin{equation}
e^{{T} \nu_k}=1\, , \qquad k=1,\dots,n\, .
\label{percons-diag}
\end{equation}
For real ${T}$, the eigenvalues $\nu_k$ has to be pure imaginary $\nu_k \equiv i \lambda_k$ and such that
\begin{equation}
{T} \lambda_k=2\pi m_k, \qquad k=1,\dots,n\, , 
\label{ratper}
\end{equation}
where $m_k$ are arbitrary integers.
These conditions impose severe constraints. First $\tr A=0$ since  $A$ is a real matrix and 
\begin{equation}
\det A=i^n \prod_{k=1}^n \lambda_k
\end{equation}
then the dimension of $A$ has to be even $n=2,4,6,\dots$.  
The eigenvalues appear in pairs $\pm i \lambda_k$.\par

Then the condition (\ref{ratper}) implies that
\begin{equation}
\frac{\lambda_k}{\lambda_l}= \frac{m_k}{m_l} \, , \qquad k,l=1,\dots,n\, , 
\end{equation}
and, consequently 
\begin{equation}
\lambda_k=\lambda \frac{p_k}{q_k}\, ,\qquad k=1,\dots,n\, ,
\label{rateigen}
\end{equation}
where $\lambda$ is real and $p_k,q_k$ are integers. \par

Using (\ref{rateigen}), one rewrites (\ref{ratper}) as
\begin{equation}
{T} \lambda \frac{p_k}{q_k}= {T} \lambda  \frac{q_1 \dots q_{k-1} p_k q_{k+1} \dots q_{n}}{q_1 \dots q_n} =2 \pi m_k \, , \qquad k=1, \dots,n\, .
\label{ratpergen}
\end{equation}
One immediately concludes that (\ref{ratpergen}) is satisfied with $\frac{{T} \lambda}{ q_1 \dots q_n}=2 \pi$ and the corresponding 
$m_k=q_1 \dots q_{k-1} p_k q_{k+1} \dots q_{n}$.\par

Hence, the condition (\ref{percons}) is satisfied for the diagonalizable matrix $A$ with eigenvalues $i \lambda \frac{p_k}{q_k}$, $k=1,\dots,n$ and
\begin{equation}
{T}=\frac{2 \pi }{\lambda} \prod_{k=1}^n q_k\, .
\label{period}
\end{equation}

Particular subclass of such matrices $A$ corresponds to the case $p_k=q_k$\, , $k=1,\dots,n$. In this case $A^2=-\lambda^2 \mathbb{I}_n$ and
\begin{equation}
e^{tA} =\cos(\lambda t) \mathbb{I}_n + \frac{1}{\lambda} \sin(\lambda t)  A\, .
\end{equation}
 Is is noted that the matrix $A$ cannot be symmetric. \par
 
 The reconstruction of the matrix $A$ is straightforward. 
 Indeed, the reality of the matrix $A$ (i.e. $A=\overline{A}$) and the condition $\overline{{A_d}}=-{A_d}$  lead to the condition
 \begin{equation}
 {A_d} B^{-1} \overline{B}+ B^{-1} \overline{B} {A_d}=0\, ,
 \end{equation}
or, equivalently,
 \begin{equation}
 (\lambda_k+\lambda_l)C_{kl}=0\, \qquad k,l=1,\dots,n\, ,
 \label{reconstep}
 \end{equation}
where $C=B^{-1} \overline{B}$. Choosing any $C$ obeying (\ref{reconstep}), one gets the constraint for the matrix $B$, namely
\begin{equation}
\overline{B}=BC\, ,
\end{equation} 
and, finally, the matrix $A$.\par

For $n=2$ one has ($\mu_1=i \lambda$, $\mu_2=-i \lambda$)
\begin{equation}
{A_d}=i \lambda \begin{pmatrix} 1& 0\\ 0 & -1 \end{pmatrix} \qquad
C=\begin{pmatrix} 0& c\\ 1/ \overline{c} & 0 \end{pmatrix} \qquad
B=\begin{pmatrix} a& \overline{c}\,  \overline{a} \\ b& \overline{c} \overline{b}  \end{pmatrix} \qquad
\end{equation}
where $a$ and $b$ are arbitrary complex numbers
and consequently,
\begin{equation}
A=\lambda \begin{pmatrix} A_{11}& A_{12}\\ A_{21} & -A_{11} \end{pmatrix}=
\frac{i\lambda}{a \overline{b} -\overline{a} b  } 
\begin{pmatrix} a \overline{b}+ \overline{a} b & 2 \overline{a} \overline{b} \overline{c}\\ -2 a b/ \overline{c} & -(a \overline{b}+ \overline{a} b) \end{pmatrix} 
\label{2D-permat}
\end{equation}
and $A_{11}^2+A_{12}A_{21}+1=0$. It noted that in this simplest $n=2$ case the matrix $A$  (\ref{2D-permat}) is immediately obtained from the conditions 
$\det A=-\lambda^2$ and $\tr A=0$.\par

For $n=4,6,8,\dots$ the calculations are straightforward, but cumbersome. At $n=4$ two simple examples of matrix $A$ are given by
\begin{equation}
A=\lambda \begin{pmatrix} 0& 1& 0& 0 \\ -1&0&0&0 \\0&0&0&1\\ 0&0&-1&0\end{pmatrix}\, , \qquad
A=\lambda \begin{pmatrix} 0& 0&1& 0 \\ 0&0&0&1 \\-1&0&0&0\\ 0&-1&0&0\end{pmatrix}\, .
\end{equation}
The inspection of the formulae (\ref{implsol-HElin}) and (\ref{hodo-HElin}) shows that for nondegenerate matrix $A$ with periodic matrix $e^{tA}$
the corresponding solutions of the equation (\ref{HEeq-lins-intro}) are periodic in time $\uu(t+{T},\ux)=\uu(t,\ux)$ where the period ${T}$ 
(\ref{period})depends on the choice of the matrix $A$.\par

It is observed that due to the periodicity of the matrix $e^{tA}$, the equation (\ref{bu-lincond}) is periodic in time too. Hence, the  corresponding blow-up
hypersurface is composed by an infinite number of sheets. \par

It is noted that in the case $\det A \neq 0$, the condition (\ref{percons}) is also sufficient for the periodicity of $\uu(t,\ux)$. As we will see in the next sections for degenerate matrix $A$ the situation is quite different.\par

Finally we note that equation (\ref{HEeq-lins-intro}) in even dimensions may have particular periodic solutions in the case $\mathbf{g} \neq 0$. For example,
independent on $\ux$ solutions (\ref{sol-cq1-kost}) and (\ref{sol-cq2-kost}) with $A$ obeying  (\ref{percons}) and (\ref{rateigen})
are periodic in time also when $\mathbf{g} \neq 0$

General analysis of the existence of periodic solutions of the Euler equation and their possible applications in physics are of interest.

\section{Two dimensional  Euler equation}
\label{sec-2dCoriolis}
In this section we will consider the two-dimensional equation (\ref{HE-lins}) with $\mathbf{g}=\mathbf{0}$  and particular choices of the matrix $A$, namely
\begin{eqnarray}
A&=& \omega \begin{pmatrix} 0&1 \\ -1&0 \end{pmatrix}\, , \qquad \omega \in \mathbb{R} \label{2D-g0-Cor}\\
A&=& a \begin{pmatrix} 1&0 \\ 0&-1 \end{pmatrix} \, , \qquad a \in \mathbb{R} \label{2D-g0-diagdeg}\\
A&=&  \begin{pmatrix} a_1&0 \\ 0&a_2 \end{pmatrix} \, , \qquad a_1,a_2 \in \mathbb{R}\, , \quad a_1\neq a_2\, . \label{2D-g0-diag} 
\end{eqnarray}
\subsection{Case $A=\omega \begin{pmatrix} 0&1 \\ -1&0 \end{pmatrix}$}
\label{subsec-2dC}
This case corresponds to the external force $\mathbf{F}_i= \rho (g_i+\omega \sum_{j=1}^2 \epsilon_{ij }u_j)$, $i=1,2$ (with $\epsilon_{11}=\epsilon_{22}=0$
$\epsilon_{12}=-\epsilon_{21}=1$) where the second term represents the Coriolis force $\mathbf{F}=-\rho \boldsymbol{\omega} \times \uu$ in two spatial 
dimensions. 
The matrix $A$ representing the two-dimensional Coriolis force is the particular instance of the matrix (\ref{2D-permat}) with $A_{11}=0$,  $A_{12}=-A_{21}=1$ and $\lambda={T}$. \par

Since in this case
\begin{equation}
e^{tA}= \begin{pmatrix} \cos(t \omega )&\sin(t \omega ) \\ -\sin(t \omega )& \cos(t \omega ) \end{pmatrix}
\end{equation}
one has
\begin{equation}
\begin{pmatrix}  N_1\\N_2     \end{pmatrix} =\begin{pmatrix}   x_1\\x_2    \end{pmatrix} 
+ \frac{1}{\omega^2}
\begin{pmatrix}  1-\cos(\omega t) & \sin(\omega t) \\ - \sin(\omega t) &   1-\cos(\omega t)  \end{pmatrix} 
\begin{pmatrix} g_1+\omega u_2\\ g_2-\omega u_1      \end{pmatrix} 
+ \frac{t}{\omega} \begin{pmatrix} -g_2\\g_1      \end{pmatrix}
\end{equation}
and
\begin{equation}
\begin{pmatrix}  M_1\\M_2     \end{pmatrix} = \frac{1}{\omega}
\begin{pmatrix}  -\sin(\omega t) & -\cos(\omega t) \\ \cos(\omega t) &   -\sin(\omega t)  \end{pmatrix} 
\begin{pmatrix} g_1+\omega u_2\\ g_2-\omega u_1      \end{pmatrix} 
+ \frac{t}{\omega} \begin{pmatrix} g_2\\-g_1      \end{pmatrix}
\end{equation}
Consequently the hodograph equations (\ref{hodo-HElin}) assume the form
\begin{equation}
\begin{pmatrix}   x_1\\x_2    \end{pmatrix} 
+ \frac{1}{\omega^2}
\begin{pmatrix}  1-\cos(\omega t) & \sin(\omega t) \\ - \sin(\omega t) &   1-\cos(\omega t)  \end{pmatrix} 
\begin{pmatrix} g_1+\omega u_2\\ g_2-\omega u_1      \end{pmatrix} 
+ \frac{t}{\omega} \begin{pmatrix} -g_2\\g_1      \end{pmatrix}=
\begin{pmatrix}  \phi_1(M_1,M_2)\\ \phi_2(M_1,M_2)     \end{pmatrix} \, .
\label{hodo-Cor2D}
\end{equation}
It is observed that in the case $\mathbf{g}=0$ the hodograph equations (\ref{hodo-Cor2D}) provides us with solutions periodic in time $t$ 
(up to blowups);  namely $\uu(t+2 \pi \omega^{-1},\ux)=\uu(t,\ux)$. It is noted that the case $\mathbf{g}=0$ and periodicity of the corresponding solutions
have been discussed earlier in the paper \cite{FKM94} using the Lagrangian variables approach.\par

In this case ($\mathbf{g}=0$) the  variables $\mathbf{M}$ are the variables $\uu$ rotated by the angle
$-t \omega$
\begin{equation}
\begin{pmatrix}  M_1\\M_2     \end{pmatrix} = 
\begin{pmatrix}  \cos(\omega t) & -\sin(\omega t) \\ \sin(\omega t) &   \cos(\omega t)  \end{pmatrix} 
\begin{pmatrix}  u_1\\ u_2      \end{pmatrix}\, .
\end{equation}
In the variables $M_1$ and $M_2$ the hodograph equation (\ref{hodo-Cor2D}) are given by ($\mathbf{g}=0$)
\begin{equation}
\begin{pmatrix}   x_1\\x_2    \end{pmatrix} 
+ \frac{1}{\omega}
\begin{pmatrix}  -\sin(\omega t) & \cos(\omega t)-1 \\ 1-\cos(\omega t)&- \sin(\omega t)     \end{pmatrix} 
\begin{pmatrix} M_1\\ M_2    \end{pmatrix} 
=
\begin{pmatrix}  \phi_1(M_1,M_2)\\ \phi_2(M_1,M_2)     \end{pmatrix} \, .
\label{hodo-Cor2D-M}
\end{equation}
The simplest solution of this equation corresponds to the function $\phi_i=\sum_{k=1}^2 R_{ik} M_k$, $i=1,2$ where $R_{ik}$ is a constant matrix
related to the initial data $u_i(t=0,\ux)=\sum_{k=1}^2 (R^{-1})_{ik} x_k$, $i=1,2$. With such a choice one gets the following solution
\begin{equation}
\begin{pmatrix}   u_1\\u_2    \end{pmatrix} =e^{tA} \Pi^{-1} \begin{pmatrix}   x_1 \\ x_2    \end{pmatrix}
\label{solCor2D-linvel0}
\end{equation}
where $\Pi^{-1}$ is the inverse of
\begin{equation}
\Pi=\begin{pmatrix} R_{11}+\frac{1}{\omega} \sin(\omega t) &   R_{12}+\frac{1}{\omega} (1-\cos(\omega t)) \\
R_{21}-\frac{1}{\omega} (1-\cos(\omega t)) & R_{22}+\frac{1}{\omega} \sin(\omega t) \end{pmatrix}
\end{equation}
assuming that $\det \Pi \neq 0$.

In the general case $\mathbf{g}\neq 0$, the blowup surface is given by the equation 
\begin{equation}
a \sin(t \omega)  + b \cos(t \omega) +c=0\, ,
\label{busur-Cor2Dg}
\end{equation}
where 
\begin{equation}
\begin{split}
a=&\omega \left(  \pp{\phi_1}{M_1} + \pp{\phi_2}{M_2} \right)\, , \\
b=& \omega  \left(  \pp{\phi_2}{M_1} -\pp{\phi_1}{M_2} \right)-2   \\
c=&-b + \omega^2 \left(  \pp{\phi_1}{M_1} \pp{\phi_2}{M_2} - \pp{\phi_1}{M_2}\pp{\phi_2}{M_1} \right) \, ,
\end{split}
\end{equation}
Equation (\ref{busur-Cor2Dg}) implies that
\begin{equation}
{t_\pm}_k= \frac{1}{\omega}\arcsin\left(\frac{-ac \pm |b| \sqrt{a^2+b^2-c^2}}{a^2+b^2}\right) + \frac{2 \pi}{\omega} k\, , \qquad k \in \mathbb{Z}\, .
\label{bucond-fun}
\end{equation}
So if 
\begin{equation}
\begin{split}
a^2+b^2&>c^2\, ,\\
a^2+b^2&> |-ac +\pm  |b| \sqrt{a^2+b^2-c^2}|\, ,
\end{split}
\label{condCor-2D}
\end{equation}
the blow-up surface $\Gamma$ has two branches, each of the consists of and infinite number of sheets.\par

In the particular case $a=0$, the conditions (\ref{condCor-2D}) are satisfied if 
\begin{equation}
\left\vert\frac{b}{c}\right\vert>1\, ,
\label{condCor-2D-part}
\end{equation}
and
\begin{equation}
{t_\pm}_k =\pm \frac{1}{\omega}\arcsin\left(  \sqrt{1-\frac{c^2}{b^2}}\right) + \frac{2 \pi}{\omega} k\, \qquad k \in \mathbb{Z}\, .
\end{equation}

If the initial data (and then the function $\boldsymbol{\phi}(\ux)$) are such that the conditions (\ref{condCor-2D}) are not satisfied,
the the equation (\ref{HEeq-lins-intro}) does not exhibit blowups. \par

For the solution (\ref{solCor2D-linvel0}) with $R_{11}+R_{22}=0$ one has
\begin{equation}
a=0\, , \qquad b=-2-\omega (R_{12}-R_{21})\, , \qquad c=-b - \omega^2 (R_{11}^2+R_{12}R_{21})\, .
\end{equation}
So, if
\begin{equation}
\left\vert -1+\frac{\omega^2 (R_{11}^2+R_{12}R_{21})}{2+\omega (R_{12}-R_{21})}  \right\vert <1
\end{equation}
the derivatives of the solution (\ref{solCor2D-linvel0}) blow-up  at the times 
\begin{equation}
{t_\pm}_k=\pm \frac{1}{\omega}\arcsin\left(  \sqrt{1-\left(1-\frac{\omega^2 (R_{11}^2+R_{12}R_{21})}{2+\omega (R_{12}-R_{21})}\right)^2}\right) + \frac{2 \pi}{\omega} k\, \qquad k \in \mathbb{Z}\, .
\label{tcat-Cor2D-linvel0}
\end{equation}
\par
We remark that in this case, the first positive blowup time ${t_-}_1$ is smaller than the period of the $2 \pi/\omega$.
The solution (\ref{solCor2D-linvel0}) has a rather special property: indeed, from (\ref{solCor2D-linvel0})  one as
\begin{equation}
\pp{u_i}{x_k}= \left( e^{tA} \Pi^{-1}\right)_{ik}\, , \qquad i,k=1,2\, .
\end{equation}
Using (\ref{der-lin}), one concludes that
\begin{equation}
e^{tA} \Pi^{-1}=K^{-1}
\end{equation}
and, hence, (\ref{solCor2D-linvel0})  becomes
\begin{equation}
u_i=\sum_{j=1}^2 K_{ij}x_j\, , \qquad i=1,2\, .
\end{equation}
So, the solutions (\ref{solCor2D-linvel0})  blow-up at times ${t_\pm}_k$ (\ref{tcat-Cor2D-linvel0}) together with its derivatives.
On the other hand, if
\begin{equation}
\left\vert -1+\frac{\omega^2 (R_{11}^2+R_{12}R_{21})}{2+\omega (R_{12}-R_{21})}  \right\vert >1
\end{equation}
The solution (\ref{solCor2D-linvel0}) is completely regular and periodic in time.\par

Finally, for solutions of the initial data (\ref{2D-exe-eps-diag}), one has 
\begin{equation}
\begin{split}
a=&\frac{\omega}{\epsilon^2-1} \frac{2-M_1^2-M_2^2}{(1-M_1^2)(1-M_2^2)}\, , \\
b=&-2+\frac{\epsilon \omega}{\epsilon^2-1} \frac{M_1^2-M_2^2}{(1-M_1^2)(1-M_2^2)}\, , \\
c=&-b-\frac{\omega}{\epsilon^2-1} \frac{1}{(1-M_1^2)(1-M_2^2)}\, .\\
\end{split}
\end{equation}

In the appendix \ref{app-ExpCor} we find the catastrophe time for Gaussian initial data and we compare it qualitatively with the numerical solution.
\subsection{Case $A=a \begin{pmatrix} 1&0 \\ 0&-1 \end{pmatrix}$}
 In the case (\ref{2D-g0-diagdeg}), defining $\tau=a^{-1}(e^{at-1})$ and $M_1=u_1 e^{at}$, $M_2=u_2 e^{-at}$ (see (\ref{cq-u0})), we obtain
 \begin{equation}
 x_1-M_1 \tau=\phi_1(M_1,M_2)\, , \qquad x_2-M_2 \frac{\tau}{1+a \tau}=\phi_2(M_1,M_2)\, .
 \end{equation}
  The blow-up surface $\Gamma$ is defined by the equation
\begin{equation}
\left( 1+ a \pp{\phi_2}{M_2}\right)\tau^2 +\left( \pp{\phi_2}{M_2}+\pp{\phi_1}{M_1}+ a Y \right) \tau +Y =0\, .
\end{equation}
where $Y \equiv \pp{\phi_2}{M_2}\pp{\phi_1}{M_1}-\pp{\phi_1}{M_2}\pp{\phi_2}{M_1}$. The roots of this equation are
\begin{equation}
\tau_\pm =\frac{1}{\left( 1+ a \pp{\phi_2}{M_2}\right)}
 \left( -  \pp{\phi_2}{M_2}-\pp{\phi_1}{M_1} +a Y \pm 
  \sqrt{\left( \pp{\phi_2}{M_2}+\pp{\phi_1}{M_1}+ a Y \right) ^2 -4Y\left( 1+ a \pp{\phi_2}{M_2}\right) }
  \right)
\end{equation}
For solutions with initial data (\ref{2D-exe-eps-diag}) one has
\begin{equation}
\tau_\pm=-\frac{ a+{M_1}^2+{M_2}^2-2  \pm \sqrt{a^2-2 a \left({M_1}^2-{M_2}^2\right)+{M_1}^2 \left({M_2}^2 \left(4 \epsilon ^2-2\right)-4 \epsilon
   ^2\right)+{M_1}^4-4 {M_2}^2 \epsilon ^2+{M_2}^4+4 \epsilon ^2}}{2 \left({M_1}^2-1\right)
   \left(a-\left({M_2}^2-1\right) \left(\epsilon ^2-1\right)\right)}
\end{equation}
\subsection{Case $A=\omega \begin{pmatrix} a_1&0 \\ 0&a_2 \end{pmatrix}$}
In this case the blowup surface is defined by the equation 
\begin{equation}
e^{t(a_1+a_2)} + e^{t a_1} K_1+ e^{t a_2} K_2+  K_3=0
\label{bu-surf-diag}
\end{equation}
where
\begin{equation}
\begin{split}
K_1=& a_2 \pp{\phi_2}{M_2}-1\, , \\
K_2=& a_1 \pp{\phi_1}{M_1}-1\, , \\
K_3=&-a_1 \pp{\phi_1}{M_1}-a_2 \pp{\phi_2}{M_2}+a_1a_2 \left(  \pp{\phi_1}{M_1} \pp{\phi_2}{M_2}- \pp{\phi_1}{M_2} \pp{\phi_2}{M_1} \right)\, .
\end{split}
\end{equation}
If the ratio  $a_1/a_2\equiv p/q$ is a rational number, the equation (\ref{bu-surf-diag}) is equivalent to the polynomial in $\tau \equiv e^{\frac{ta_2}{q}}=e^{\frac{ta_1}{p}}$ 
\begin{equation}
\tau^{p+q} + K_1 \tau^{p}+ K_2 \tau^{q}+K_3=0\, .
\end{equation}
For irrational $a_1/a_2$ the analysis is more involved. However, the calculation of the minimal or maximal values of $t$ for which 
$\pp{t}{M_1}\Big{\vert}_{t=t^*}=\pp{t}{M_2}\Big{\vert}_{t=t^*}=0$ is, in principle, a rather straightforward procedure.\par

Indeed, differentiating (\ref{bu-surf-diag}) with respect to $M_1$ and $M_2$ and evaluating the result $t=t^*$, one gets
\begin{equation}
\begin{split}
& e^{t^* a_1} \pp{K_1}{M_1}+e^{t^* a_2} \pp{K_2}{M_1}+ \pp{K_3}{M_1}=0\, , \\
& e^{t^* a_1} \pp{K_1}{M_2}+e^{t^* a_2} \pp{K_2}{M_2}+ \pp{K_3}{M_2}=0\, .
\end{split}
\end{equation}
This system generically implies that 
\begin{equation}
e^{t^* a_1}=f_1(M_1,M_2)\, , \qquad e^{t^* a_2}=f_2(M_1,M_2)\, ,
\label{min-t-nonrat}
\end{equation}
where $f_1(M_1,M_2)$ and $f_2(M_1,M_2)$ are certain functions subject to the constraint
\begin{equation}
\left( f_1(M_1,M_2)\ \right)^{a_2} = \left( f_2(M_1,M_2)\ \right)^{a_1}\, .
\end{equation}
Substituting  (\ref{min-t-nonrat}) into (\ref{bu-surf-diag}), one obtains  another constraint for $M_1$ and $M_2$. Thus, $t$ has minimum or
maximum in a point $M_1^*$ and $M_2^*$ and $t^*$is easily obtained from the formulae (\ref{min-t-nonrat}).

\section{The case of degenerate matrix $A$}
\label{sec-degenA}
In the previous sections we discussed the case of $\det(A)\neq 0$. If $\det(A) = 0$, the formulae presented in section \ref{sec-linsource} 
have to be modified. \par

Let the rank of the matrix $A$ be equal to $r<n$. In this case 	there are $n-r$ linearly independent vectors $\mathbf{L}^{(\alpha)}$, $ \alpha=1,\dots, n-r$
such that (see e.g. \cite{Gel89})
\begin{equation}
\sum_{i=1}^n {L}^{(\alpha)}_i A_{ik}=0\, \qquad  \alpha=1,\dots, n-r\, .
\end{equation}
The structure of the equation (\ref{HEeq-lins-intro}) and (\ref{linvel-cq1}), (\ref{linvel-cq2}) clearly 
indicate that the following linear combinations
\begin{equation}
y_\alpha \equiv \sum_{i=1}^n {L}^{(\alpha)}_i x_i= \mathbf{L}^{(\alpha)} \cdot \ux\, , \qquad 
v_\alpha \equiv \sum_{i=1}^n {L}^{(\alpha)}_i u_i=\mathbf{L}^{(\alpha)} \cdot \uu \, , \qquad 
f_\alpha \equiv \sum_{i=1}^n {L}^{(\alpha)}_i g_i=\mathbf{L}^{(\alpha)} \cdot \mathbf{g} \, , \qquad \alpha=1,\dots,n-r\, ,
\label{kerA-var}
\end{equation}
have a special role.
For instance, one has
\begin{equation}
\pp{v_\alpha}{t}+ \sum_{k}u_k \pp{v_\alpha}{x_k}=f_\alpha\, , \qquad 
I^{(1)}_\alpha=\sum_{k=1}^{n} L_{k}^{(\alpha)} I^{(1)}_k = v_\alpha-f_\alpha t\, , \qquad   
I^{(2)}_\alpha=\sum_{k=1}^n L_{k}^{(\alpha)} I^{(2)}_k=f_\alpha , \qquad  \alpha=1, \dots,n-r\, .
\end{equation}
To exploit this fact we introduce  new variables
\begin{equation}
y_a \equiv \sum_{i=1}^n {L}^{(a)}_i x_i= \mathbf{L}^{(a)} \cdot \ux\, , \qquad 
v_a \equiv \sum_{i=1}^n {L}^{(a)}_i u_i=\mathbf{L}^{(a)} \cdot \uu \, , \qquad 
f_a \equiv \sum_{i=1}^n {L}^{(a)}_i g_i=\mathbf{L}^{(a)} \cdot \mathbf{g}\, , \qquad a=1,\dots,n \, .
\label{new-kerA-var}
\end{equation}
where the vectors $\mathbf{{L}^{(i)}}$, $i=n-r+1\dots,n$ are complementary to the vectors  $\mathbf{{L}^{(i)}}$, $i=1,\dots,n-r$ and all of them
form an orthogonal basis of $\mathbb{R}^n$. \par

In these variables, the equation (\ref{HEeq-lins-intro}) assumes the form
\begin{equation}
\begin{split}
&\pp{v_\alpha}{t}+\sum_{k=1}^n v_k \pp{v_\alpha}{y_k}=f_\alpha\, \qquad \alpha=1,\dots,n-r\, , \\
&\pp{v_\beta}{t}+\sum_{k=1}^n v_k \pp{v_\beta}{y_k}=f_\beta+ \sum_{k=1}^n B_{\beta k} v_k \ \, \qquad \beta=n-r+1,\dots,n\, , \\
\end{split}
\label{HEeq-lins-intro-kerA}
\end{equation}
where
\begin{equation}
\begin{split}
&B_{\beta i}= \sum_{k,l=0}^n L^{(\beta)}_{k} A_{kl} P^{(l)}_{i} \, \qquad \beta=n-r+1,\dots,n\, , \qquad i=1,\dots,n\, ,\\
\mathrm{and}  \qquad & \sum_{k=1}^n  P^{(i)}_{k} L^{(k)}_{l}=\delta_{il}\, ,\qquad 
\sum_{k=1}^n  L^{(i)}_{k} P^{(k)}_{l} =\delta_{il}\, ,\qquad i,l=1,\dots,n\, .
\end{split}
\label{Bmat}
\end{equation}
Equation (\ref{eik-lins}) becomes
\begin{equation}
\pp{S_i}{t}+ \sum_{k=1}^n v_k \pp{S_i}{y_k}+ \sum_{\alpha=1}^{n-r} f_\alpha \pp{S_i}{y_\alpha}+
\sum_{\beta={n-k+1}}^n \left(f_\beta+ \sum_{k=1}^n B_{\beta k} v_k\right) \pp{S_i}{v_\beta}=0\, ,
\label{eik-lins-kerA}
\end{equation}
and equations for characteristics are
\begin{equation}
\frac{\D y_i}{\D t}=v_i\, , \qquad \frac{\D v_\alpha}{\D t}=f_\alpha \, , \quad \alpha=1,\dots, n-r,
 \qquad \frac{\D v_\beta}{\D t}=f_\beta+ \sum_{k=1}^n B_{\beta k} v_k \, , \quad \beta=n-r+1,\dots, n.
 \label{char-kerA}
\end{equation}
Integrals (\ref{linvel-cq1}) and  (\ref{linvel-cq2}) are given by
\begin{equation}
\begin{split}
I^{(1)}_\alpha &= v_\alpha-f_\alpha t\, , \qquad   I^{(2)}_\alpha=f_\alpha , \qquad  \alpha=1, \dots,n-r\, , \\
I^{(1)}_\beta &=v_\beta-f_\beta t-\sum_{k=1}^n B_{\beta k} y_k \, , \qquad
I^{(2)}_\beta = \sum_{k=1}^n C_{\beta k} f_k+\sum_{k=1}^n D_{\beta k} v_k\, ,
\qquad \beta=n-r+1, \dots,n\, ,
\end{split}
\end{equation}
where
\begin{equation}
C_{\beta i }(t)=\sum_{k,l=1}^n L^{(\beta)}_k   \left( e^{-tA}\right)_{kl} P^{(l)}_i  \,, \qquad 
D_{\beta i }(t)=\sum_{k,l=1}^n L^{(\beta)}_k   \left( A e^{-tA}\right)_{kl} P^{(l)}_i  \,, \qquad 
 \beta=n-r+1,\dots,n \, , \quad i=1,\dots,n\, .
 \label{CDmat}
\end{equation}
Note that $C_{\beta i }(t=0)=\delta_{\beta i}$ and $D_{\beta i }(t=0)=B_{\beta i}$ with $ \beta=n-r+1,\dots,n $ and $ i=1,\dots,n$.\par

Analogs of the integrals $\mathbf{M}$ and $\mathbf{N}$ (\ref{cq-u0}), (\ref{lin-cq-xu}) are of the form 
\begin{equation}
\begin{split}
M_\alpha =&v_\alpha - f_\alpha t\, , \qquad N_\alpha=y_\alpha-v_\alpha t + \frac{1}{2}f_\alpha t^2 \, , \qquad \alpha=1, \dots,n-r\, \\
M_\beta=&\sum_{\gamma=n-r+1}^n \left( \tilde{B}^{-1}\right)_{\beta \gamma}
\left[  \sum_{k=1}^n (D_{\gamma k} v_k + C_{\gamma k} f_k) - f_{\gamma} +  \sum_{\alpha=1}^{n-r} D_{\gamma \alpha} v_\alpha-
  \sum_{\alpha=1}^{n-r} B_{\gamma \alpha} (v_\alpha -  f_\alpha t) 
\right]\, , \qquad \beta=n-r+1, \dots, n, \\
N_\beta=&y_\beta + \sum_{\gamma=n-r+1}^n \left( \tilde{B}^{-1}\right)_{\beta \gamma}
\left[ M_{\gamma}-v_\gamma+f_\gamma t + \sum_{\alpha=1}^{n-r} B_{\gamma \alpha}\left(v_\alpha t - \frac{1}{2}f_\alpha t^2 \right)
\right]\, , \qquad \beta=n-r+1, \dots, n,  
\end{split}
\label{cq12-lins-kerA}
\end{equation}
where $\tilde{B}$ is the $r\times r$ matrix with elements $B_{\beta \gamma}$, $\beta \gamma=n-r+1,\dots,n$ (assuming $\det(\tilde{B}) \neq 0$).\par

At $t=0$ one has
\begin{equation}
\mathbf{M}(t=0)=\mathbf{v} \, ,\qquad
\mathbf{N}(t=0)=\mathbf{y} \, .
\end{equation}

Resolving the basic equation (\ref{eik-lin-xu0}) w.r.t. $\mathbf{N}$, one gets the hodograph equations
\begin{equation}
\begin{split}
&y_\alpha-v_\alpha t +\frac{1}{2} f_\alpha t^2=\phi_\alpha(\mathbf{M})\, , \qquad \alpha=1, \dots, n-r\\
&y_\beta+ 
\sum_{\gamma=n-r+1}^n \left( \tilde{B}^{-1}\right)_{\beta \gamma}
\left[ M_{ \gamma}-v_\gamma+f_\gamma t + \sum_{\alpha=1}^{n-r} B_{\gamma \alpha}\left(v_\alpha t -  \frac{1}{2}f_\alpha t^2 \right)
\right]
=\phi_\beta(\mathbf{M})\, , \qquad \beta=n-r+1, \dots, n\, .
\end{split}
\label{HEeq-lins-hodo-kerA}
\end{equation}
Is is straight forward to check that these hodograph equations provide us with solutions $\mathbf{v}$ of the equation (\ref{HEeq-lins-intro-kerA}), i.e.  
(\ref{HEeq-lins-intro}) in the case $\det A=0$. \par

Note that at $t=0$ 
\begin{equation}
\mathbf{y}=\boldsymbol{\phi}(\mathbf{v}(t=0))\, .
\end{equation}
So, the functions $\boldsymbol{\phi}$ are inverse to the initial data $\mathbf{v}(t=0,\mathbf{y})$. Note also that the hodograph equation 
(\ref{HEeq-lins-hodo-kerA}) can be rewritten in terms of variables $\mathbf{M}$ that, as in the generic case, simplifies the analysis of the 
blow-ups conditions.\par

Note that in this case with degenerate matrix $A$, the matrix  $e^{tA}$  is periodic in time while the hodograph equation
(\ref{HEeq-lins-hodo-kerA}) and consequently solutions $\uu$ are not, even when $\mathbf{g}=0$.

\section{Coriolis force in 3D}
\label{sec-3DCor}
Three dimensional pressureless Euler equation is a notable example of equation (\ref{HEeq-lins-intro}) with degenerate matrix $A$. \par
For the Coriolis force $\mathbf{F}=\rho(\mathbf{g}-\boldsymbol{\omega} \times \uu)$ where $\boldsymbol{\omega}$ is the 
of an angular velocity vector, the matrix A is of the form
\begin{equation}
A=\begin{pmatrix}
0& \omega_3 & -\omega_2 \\
-\omega_3 & 0 &\omega_1 \\
\omega_2 & \omega_1&0 
\end{pmatrix}
\label{matA-Cor3D}
\end{equation}
The matrix (\ref{matA-Cor3D}) has generically rank two and the normalized vector $\mathbf{L^{(1)}}$ generating the left kernel of A is
\begin{equation}
\mathbf{L^{(1)}}=\frac{1}{|\boldsymbol{\omega}|}\boldsymbol{\omega}\, .
\end{equation}
Vectors $\mathbf{L^{(2)}}$ and $\mathbf{L^{(3)}}$ (forming with $\mathbf{L^{(1)}}$ a basis of $\mathbb{R}^3$) can be chosen as
\begin{equation}
\mathbf{L^{(2)}}=\frac{1}{\sqrt{\omega_2^2+\omega_3^2}}(0,\omega_3,-\omega_2)\, , \qquad
\mathbf{L^{(3)}}=\frac{1}{|\boldsymbol{\omega}|\sqrt{\omega_2^2+\omega_3^2}}(\omega_2^2+\omega_3^2,-\omega_1\omega_2,-\omega_1\omega_3,)\, .
\label{kerbas}
\end{equation}
Accordingly
\begin{equation}
\begin{split}
\mathbf{P^{(1)}}&= \frac{1}{|\boldsymbol{\omega}|} \left(\omega_1,0,  {\sqrt{\omega_2^2+\omega_3^2}}\right)\, ,  \\
\mathbf{P^{(2)}}&= \frac{1}{|\boldsymbol{\omega}|} \left(\omega_2,
 \frac{|\boldsymbol{\omega} |\omega_3}{\sqrt{\omega_2^2+\omega_3^2}},-\frac{\omega_1\omega_2}{\sqrt{\omega_2^2+\omega_3^2}} \right)\, , \\
\mathbf{P^{(3)}}&= \frac{1}{|\boldsymbol{\omega}|} \left(\omega_3,
- \frac{|\boldsymbol{\omega}| \omega_2}{\sqrt{\omega_2^2+\omega_3^2}},-\frac{\omega_1\omega_3}{\sqrt{\omega_2^2+\omega_3^2}} \right) \, .
\label{invkerbas}
\end{split}
\end{equation}
Note that $\det L = \det P =-1$ and $L$ and $P=L^{-1}$ are $3 \times 3$ matrices with elements $L^{(k)}_i$ and $P^{(k)}_i$ , $i,k=1,2,3$
respectively.\par

Since the matrix $A$ is skew-symmetric (then also $\sum_{j}A_{ij}L^{(1)}_j=0$, $i=1,2,3$) the matrix  $e^{tA}$
is an element of the group $SO(3)$
of rotations describing the rotation on the angle $t \omega$ around the axis $\mathbf{L^{(1)}}=\frac{\boldsymbol{\omega}}{|\boldsymbol{\omega}|}$.
The vector $\mathbf{L^{(1)}}$ is obviously invariant under such rotations $\sum_{j} (e^{tA})_{ij}L^{(1)}_j=L^{(1)}_i$, $i=1,2,3$.\par

The variables $y_1,v_1$ and $f_1$ are the projections of $\ux$, $\uu$ and $\mathbf{g}$.\par

Using (\ref{matA-Cor3D})-(\ref{invkerbas}) one obtains (see (\ref{Bmat}))
\begin{equation}
B_{21}=B_{31}=B_{22}=B_{33}=0\, , \qquad  B_{32}=-B_{23}=|\boldsymbol{\omega}|.
\end{equation}
Then calculating the matrices $C$ and $D$ (\ref{CDmat}) 
\begin{equation}
\begin{split}
&C_{11}=1\, , \qquad C_{22}=C_{33}=\cos(|\boldsymbol{\omega}|t)\, , \qquad C_{23}=-C_{32}=\sin(|\boldsymbol{\omega}|t)\, , 
\qquad  C_{12}=C_{13}=C_{21}=C_{31}=0\, .\\
&D_{22}=D_{33}=|\boldsymbol{\omega}| \sin(|\boldsymbol{\omega}|t)\, , \qquad  
D_{32}=D_{23}=|\boldsymbol{\omega}| \cos(|\boldsymbol{\omega}|t)\, , \qquad  D_{11}=D_{12}=D_{13}=D_{21}=D_{31}=0\, .\\
\end{split}
\end{equation}
one gets  explicit form of the integrals $\mathbf{M}$  and $\mathbf{N}$ and the hodograph equations (\ref{cq12-lins-kerA}) 
and (\ref{HEeq-lins-hodo-kerA}) with $r=2$. \par

In the case $\mathbf{g}=0$ they are given by 
\begin{equation}
\begin{split}
&y_1-v_1 t = \phi_1(\mathbf{M}) \, ,\\
&y_\beta+\sum_{\gamma=2,3} (\tilde{B}^{-1})_{\beta \gamma} (M_\gamma-v_\gamma)=\phi_\beta(\mathbf{M})\, , \qquad \beta=2,3\, .
\end{split}
\label{hodo-Cor3Dgen}
\end{equation}
where we used the fact that $B_{\beta 1}=0$, $\beta=1,2,3$. One immediately observes that, though the matrix $e^{tA}$ is periodic in time $t$  
with period $2 \pi / |\boldsymbol{\omega}|$, the hodograph equation (\ref{hodo-Cor3Dgen}) are not due to the presence 
of the term $v_1 t$. So, solutions provided by equations (\ref{hodo-Cor3Dgen}) are not periodic too; it is direct consequence of the degeneracy 
of the matrix $A$.\par

Special case of the equation (\ref{HEeq-lins-intro}) with $\mathbf{g}=- g \mathbf{k}$  
and $\boldsymbol{\omega} = \omega \mathbf{k}$ where $ \mathbf{k}$
is a fixed vector is of particular interest since it mimics the effect of the Coriolis force on the surface of Earth.

With the choice 
$\mathbf{k} = (0,0,1)$, the equations of motion are
\begin{equation}
\begin{split}
{u}_t&+\uu \cdot \nabla u  - \omega v=0\, , \\
{v}_t&+\uu \cdot \nabla v  + \omega u=0\, ,\\
{w}_t&+\uu \cdot \nabla w +g=0\, ,
\end{split}
\label{eqCor3D}
\end{equation}
where $\uu=(u,v,w)$ and
the matrix $A$ is of the form
\begin{equation}
A=\omega \begin{pmatrix} 0&1&0 \\ -1&0&0 \\ 0&0&0 \end{pmatrix}.
\end{equation}
Then one has
\begin{equation}
\mathbf{L^{(1)}}=\mathbf{P^{(1)}}=(0,0,1)\, , \qquad
\mathbf{L^{(2)}}=\mathbf{P^{(2)}}=(0,1,0)\, , \qquad
\mathbf{L^{(3)}}=\mathbf{P^{(3)}}=(1,0,0)\, .
\end{equation}
Note that $\mathbf{L^{(i)}}$ and  $\mathbf{P^{(i)}}$ are of the form (\ref{kerbas}) and (\ref{invkerbas}) at $\omega_1=\omega_2=0$, $\omega_3=1$.\par

The variables $\mathbf{v}$, $\mathbf{y}$, $\mathbf{f}$ are
\begin{equation}
\begin{split}
v_1&=w\, , \quad v_2=v\, , \quad v_3=u\, ,\\
y_1&=z\, , \quad y_2=y\, , \quad y_3=x\, ,\\
f_1&=-g\, , \quad f_2=0\, , \quad f_3=0\, .
\end{split}
\end{equation}
This apparently unusual choice is due to the need to adapt the  formulas of this section with the formulas of the previous one in (\ref{kerA-var}). \par

The matrix $e^{tA}$ is
\begin{equation}
e^{tA}=
\begin{pmatrix}
 \cos (t \omega) & \sin (t \omega) & 0 \\
 -\sin (t \omega) & \cos (t \omega) & 0 \\
 0 & 0 & 1 \\
\end{pmatrix}
\end{equation}
Direct calculations give
\begin{equation}
B=\omega \begin{pmatrix}
0& 0 & 0 \\
0 & 0 & -1 \\
 0 & 1 & 0 \\
\end{pmatrix} \, , \qquad
C=\begin{pmatrix}
1& 0 & 0 \\
0 &  \cos (t \omega) & \sin (t \omega) \\
 0 &  -\sin (t \omega) & \cos (t \omega)\\
\end{pmatrix}\, , \qquad
D=\omega \begin{pmatrix}
0& 0 & 0 \\
0 &  \sin (t \omega) & -\cos (t \omega) \\
 0 &  \cos (t \omega) & \sin (t \omega)\\
\end{pmatrix}\, .
\end{equation}

Consequently the integrals (\ref{cq12-lins-kerA})
\begin{equation}
\begin{split}
M_1&= w+gt \, , \\
M_2&= \sin(\omega t ) u + \cos(\omega t ) v \, ,\\
M_3&= \cos(\omega t ) u - \sin(\omega t ) v \, , 
\end{split}
\label{3DCor-M}
\end{equation}
and 
\begin{equation}
\begin{split}
N_1&= z-wt-\frac{1}{2}gt^2 \, , \\
N_2&= y+\frac{1}{\omega} \left(-(1-\cos(\omega t ) ) \, u - \sin(\omega t) \, v \right)  \, . \\
N_3&=x+\frac{1}{\omega} \left( -\sin(\omega t )\, u + (1-\cos(\omega t))\,  v \right) \, , \\
\end{split}
\label{3DCor-N}
\end{equation}
Hodograph equations (\ref{HEeq-lins-hodo-kerA}) become 
\begin{equation}
\begin{split}
&x+\frac{1}{\omega} \left( -\sin(\omega t )\, u + (1-\cos(\omega t))\,  v \right) 
= \phi_1\Big(\cos(\omega t ) u - \sin(\omega t ) v,\sin(\omega t ) u + \cos(\omega t ) v ,w+gt \Big)\, , \\
&y+\frac{1}{\omega} \left((\cos(\omega t ) -1) \, u - \sin(\omega t) \, v \right) 
= \phi_2\Big(\cos(\omega t ) u - \sin(\omega t ) v,\sin(\omega t ) u + \cos(\omega t ) v ,w+gt \Big) \, , \\
&z-wt-\frac{1}{2}gt^2 
= \phi_3\Big(\cos(\omega t ) u - \sin(\omega t ) v,\sin(\omega t ) u + \cos(\omega t ) v ,w+gt \Big)\, , 
\end{split}
\label{3DCor-hodo}
\end{equation}

In terms of the variables $\mathbf{M}$ the hodograph equations are
\begin{equation}
\begin{split}
x-\frac{M_3}{\omega} \sin(\omega t) -\frac{M_2}{\omega} (1-\cos(\omega t))=& \phi_1(M_3,M_2,M_1)\, , \\
y-\frac{M_2}{\omega} \sin(\omega t) + \frac{M_3}{\omega} (1-\cos(\omega t))=& \phi_2(M_3,M_2,M_1)\, , \\
z-M_1 t +\frac{1}{2}gt^2=& \phi_3(M_3,M_2,M_1)\,. 
\end{split}
\end{equation}
As in the generic case the variables $\mathbf{M}$ are convenient in the analysis of blowups. In particular, the blowup surface $\Gamma$ is
defined by the equation
\begin{equation}
\det\left( \pp{\phi_i}{M_k}-\mathcal{L}_{ik}\right)=0\, ,
\label{Cor3Dgeo-bus}
\end{equation}
where the matrix $\mathcal{L}$ is
\begin{equation}
\mathcal{L}=\begin{pmatrix} 
0 & -\frac{1-\cos(t \omega)}{\omega} & -\frac{\sin(t \omega)}{\omega}\\
0 & -\frac{\sin(t \omega)}{\omega}  & \frac{1-\cos(t \omega)}{\omega}\\
t&0&0\\
\end{pmatrix}
\end{equation}

The transcendental equation (\ref{Cor3Dgeo-bus}) is rather cumbersome. At $\omega t \ll 1$ it becomes a cubic equation because  $\mathcal{L}$
becomes
\begin{equation}
\mathcal{L}_{\omega t \ll 1}=\begin{pmatrix} 
0 & 0& -t \\
0 & -t& 0 \\
t&0&0
\end{pmatrix}\, .
\end{equation}

\appendix
\section{Appendix: generic 1D case}
\label{app-gen1D}
Let us first consider the solution of the one dimensional Euler equation (\ref{HE-lins}) with the initial data (\ref{condinitanh1D-g})
\begin{equation}
u_t+uu_x=Au+g\, , \qquad u(x,0)=\mu \left( 1-\tanh \left( \kappa x \right)\right)\, .
\end{equation}
 The blowup curve (\ref{bu-cond-lin-1D}) in this case is given by
 \begin{equation}
 t=\frac{1}{A} \log \left( 1+ \frac{A  \mu}{\kappa M(2\mu-M)}\right)
 \end{equation}
So, it is a real curve if 
\begin{equation}
 \frac{A \mu}{\kappa M(2\mu-M)}>-1\, .
\end{equation}
In the case
\begin{equation}
 \frac{A  \mu}{\kappa M(2\mu-M)}\leq-1\, ,
 \label{nobugen}
\end{equation}
for $M$ in the interval $0 \leq M \leq 2\mu$, solution with initial data (\ref{condinitanh1D-g}) does not exhibit blowup. Since $M(2\mu-M)>0$ and 
$\max(M(2u-M))=\mu^2$, the condition (\ref{nobugen}) is equivalent to
\begin{equation}
A <-\frac{\mu}{L}.
\end{equation}
Thus, for $\mu>0$ the solution with the initial data (\ref{condinitanh1D-g}) does not blowup if $A<0$ and $|A|>\kappa \mu$. When $\mu<0$ the corresponding condition is 
$A>\kappa |\mu|$. Note that $t_b=1/\mu$ is the extreme (minimal or maximal) blowup time for te HEE with $A=0=g$. \par

It is noted also that in the case $A<0$ the solution of the Euler equation with the initial data (\ref{condinitanh1D-g}) and $\mu<0$ stil exhibits blowups of derivatives for negative $t$.

\begin{figure}[H]
\begin{center}
\includegraphics[width=.4 \textwidth]{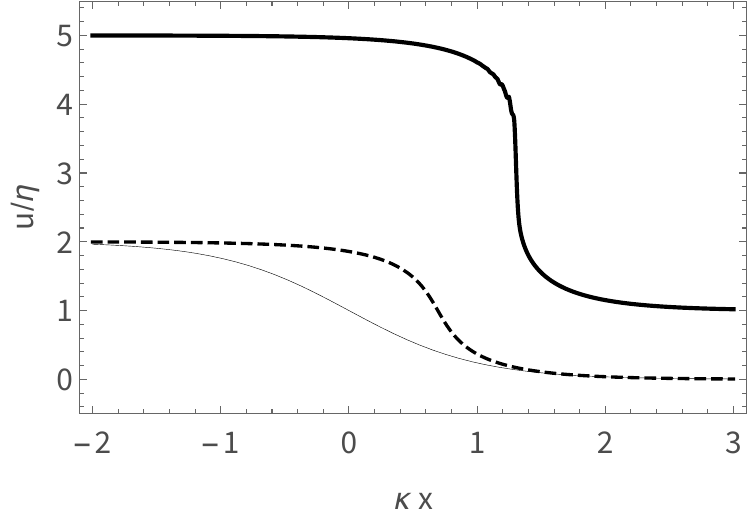}
\caption{Evolution of the initial datum $u=1-\tanh(x)$, corresponding to (\ref{condinitanh1D-g}) with $\mu=\kappa=1$. 
The thin line is the initial datum $u(x,t=0)$, the dashed line is $u(x,t_c=1)$ with $A=g=0$ while the bold line is $u(x,t_c=\log(2))$ with $A=g=1$. The 
discontinuity on the bold line at $\kappa  x \simeq 1.4$ is a numerical artifact due to the proximity to the catastrophe.}
\label{fig-cat-1D-velgrav}
\end{center}
\end{figure}

In the figure \ref{fig-cat-1D-velgrav} we show the evolution of the initial data on the class (\ref{condinitanh1D-g}). The explicit critical values are
\begin{equation}
\begin{split}
M_c=&\mu\, ,\qquad 
t_c= \frac{1}{A}\log\left( 1+\frac{A}{\kappa  \mu}\right)\, ,\qquad
u_c=\mu+\frac{A}{\kappa }+\frac{g}{\kappa \mu}\, ,\qquad\ \\
x_c=&\frac{1}{A}\log\left( 1+\frac{A}{\kappa  \mu}\right)   \left( \mu+\frac{A}{\kappa }+\frac{g}{\kappa \mu} \right) 
- \frac{g}{2A^2}\log^2\left( 1+\frac{A}{\kappa  \mu}\right)
+ \frac{1}{\kappa} \mathrm{atanh} \left( \mu+\frac{A}{\kappa }+\frac{g}{\kappa \mu} -\frac{g}{A}\log\left( 1+\frac{A}{\kappa  \mu}\right) \right) \, .\qquad
\end{split}
\end{equation}
 Other solutions of the Euler equation with external force $F=\rho(g+Au)$ have similar properties. Indeed, let 
 \begin{equation}
 u_0(x)=\eta e^{-\kappa^2 x^2}\, , \qquad \eta,\kappa >0 .
 \label{inicond-gauss-1D}
 \end{equation}
 In this case the function $\phi$ has two branches
 \begin{equation}
 \begin{split}
 \phi_+(u)&= \frac{1}{\kappa}\left(\log \left(\frac{\eta}{u} \right)\right)^{1/2}\, , \qquad  x>0\, ,\\
 \phi_-(u)&= -\frac{1}{\kappa}\left(\log \left(\frac{\eta}{u} \right)\right)^{1/2}\, , \qquad  x<0\, ,
 \end{split}
 \end{equation}
 and the hodograph equation (\ref{hodo-lin1D}) is composed by two of them, one for $x>0$ and $\phi =\phi_+$ and the second for $x<0$
 and $\phi =\phi_-$.\par
 
The blowup curve (\ref{bu-cond-lin-1D}) is given by 
\begin{equation}
t=\frac{1}{A} \log \left( 1+ \frac{\epsilon A}{2 \kappa M} \left( \log\left( \frac{\eta}{M} \right)\right)^{-1/2}  \right), \qquad 0 \leq M \leq 1\, ,
\end{equation}
with $\epsilon=\pm 1$ when $\phi=\phi_\pm$.
 This equation has no real roots if 
 \begin{equation}
 1+ \frac{\epsilon A}{2 \kappa M} \left(\log\left( \frac{\eta}{M}\right)\right)^{-1/2}<0\, ,
 \label{exp-norr}
 \end{equation}
The condition (\ref{exp-norr})
is equivalent to
\begin{equation}
\epsilon A < -\kappa \eta \left( \frac{e}{2} \right)^{-1/2}\, .
\end{equation}
As in the previous case, for $A<0$ one has the condition
\begin{equation}
\epsilon |A| > \kappa \eta \left( \frac{e}{2} \right)^{-1/2}\, .
\end{equation}
Then the branch of solution with initial condition (\ref{inicond-gauss-1D}) and $x>0$ ($\epsilon=1$) does not exhibit blowup (and therefore gradient catastrophe).  On the other hand, the branch $x<0$ ($\epsilon=-1$) still has the blow-up since in this case 
\begin{equation}
1+ \frac{\epsilon A}{2 \kappa M} \left(\log\left( \frac{\eta}{M}\right)\right)^{-1/2}>0\, .
\end{equation}
For $A>0$ one has the inverse situation. So the presence of the external force $F=\rho(\rho+Au)$ does not eliminate the blowups of the
solution with gaussian initial data (\ref{inicond-gauss-1D}).


\section{Gaussian initial data for 2D Coriolis case}
\label{app-ExpCor} 
\begin{figure}[h]
\begin{center}
\includegraphics[width=.5 \textwidth]{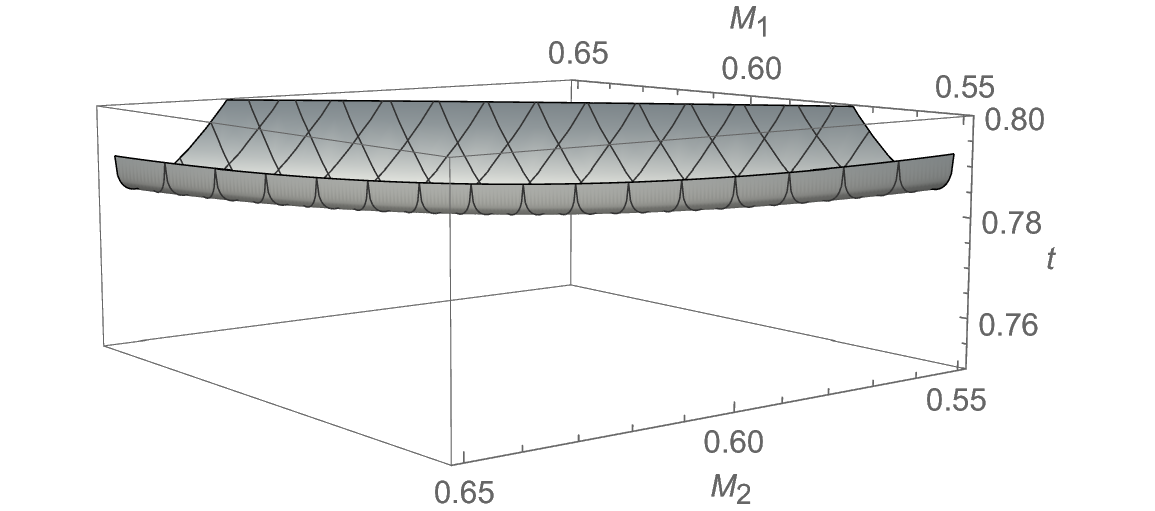}
\caption{Behavior of a branch of the blowup surface (\ref{busur-exe}) near to the smallest positive minimum.}
\label{fig-gauss-busur}
\end{center}
\end{figure}

In this section we will compute the catastrophe time and compare it with  the numerics  in the Coriolis case 
without constant force 
\begin{equation}
F_i= \sum_{j=1}^2 A_{ij} u_j\, ,\qquad A=\begin{pmatrix} 0 &1  \\ -1 &0  \end{pmatrix}.
\end{equation} 
Let us consider initial data
\begin{equation}
u_1(\ux,0)= e^{-(x^2+y^2)}\, , \qquad
u_2(\ux,0)= e^{-(x^2+2y^2)}\, , 
\label{gauss-exe}
\end{equation} 
which imply
\begin{equation}
\boldsymbol{\phi}(\mathbf{M})=\left( \pm \log \left( \frac{M_2}{{M_1}^2} \right), \pm \log \left(\frac{M_1}{M_2} \right) \right)
\end{equation}
with
\begin{equation}
\begin{pmatrix}  M_1\\M_2     \end{pmatrix} = 
\begin{pmatrix}  \cos(\omega t) & -\sin(\omega t) \\ \sin(\omega t) &   \cos(\omega t)  \end{pmatrix} 
\begin{pmatrix}  u_1\\ u_2      \end{pmatrix}\, .
\end{equation}
See subsection \ref{subsec-2dC} for details. 
The blowup surface is given by the equation  (see (\ref{catsur-intro}))
\begin{equation}
\det \left(  \left(
\begin{array}{cc}
 \sin (t) & 1-\cos (t) \\
 \cos (t)-1 & \sin (t) \\
\end{array}
\right)  + 
\left(
\begin{array}{cc}
\sin(t) -\frac{1}{{M_1}\sqrt{\log
   \left(\frac{{M_2}}{{M_1}^2}\right)}} & 
   1-\cos(t)+\frac{1}{2
   {M_2}\sqrt{\log
   \left(\frac{{M_2}}{{M_1}^2}\right)}} \\
 \cos(t)-1+\frac{1}{2 {M_1}\sqrt{\log
   \left(\frac{{M_1}}{{M_2}}\right)}} & \sin(t)-\frac{1}{2
   {M_2} \sqrt{\log
   \left(\frac{{M_1}}{{M_2}}\right)}} \\
\end{array}
\right)
\right)
\end{equation}
The blowup  surfaces $t=t(\mathbf{M})$  are given by the formula (\ref{bucond-fun})
\begin{equation}
{t_\pm}= \arcsin\left(\frac{-ac \pm |b| \sqrt{a^2+b^2-c^2}}{a^2+b^2}\right) \, ,
\label{busur-exe}
\end{equation}
where
\begin{equation}
\begin{split}
a=&-\frac{1}{M_1 \sqrt{\log
   \left(\frac{M_2}{M_1^2}\right)}}-\frac{1}{2 M_2 \sqrt{\log
   \left(\frac{M_1}{M_2}\right)}}\\
b=&-\frac{1}{2M_2 \sqrt{\log
   \left(\frac{M_2}{M_1^2}\right)}}+\frac{1}{2M_1 \sqrt{\log
   \left(\frac{M_1}{M_2}\right)}}-2 \\
c=&-b+\frac{1}{4 M_1 M_2 \sqrt{\log \left(\frac{M_1}{M_2}\right)}
   \sqrt{\log \left(\frac{M_2}{M_1^2}\right)}}   \, .
\end{split}
\end{equation}
Plotting with Mathematica the graph of a branch of the blowup surface (see fig. \ref{fig-gauss-busur}), the catastrophe time can be estimated  as $t_c \simeq 
0.78$.  

\begin{figure}[H]
\begin{center}
\includegraphics[width=.7 \textwidth]{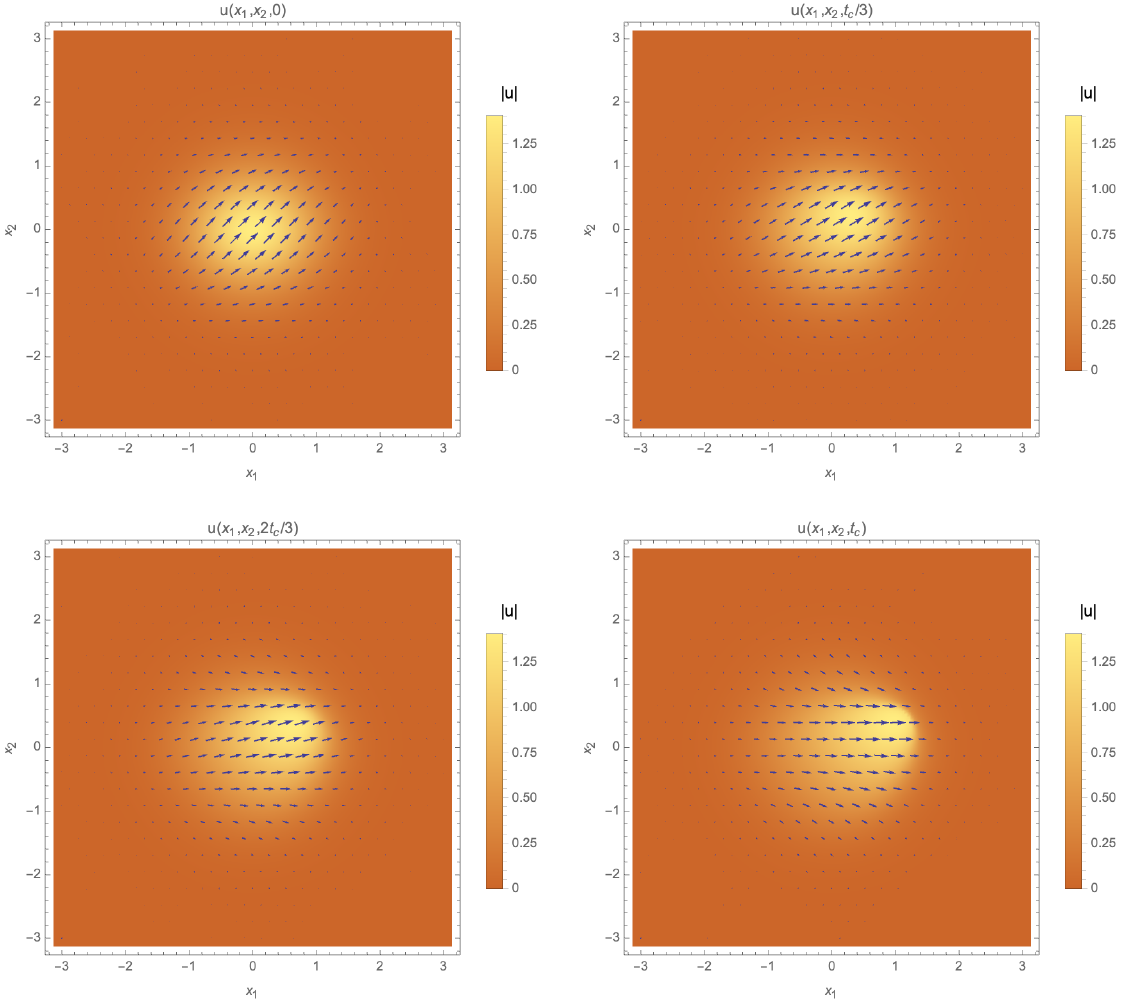}
\caption{Numerical evolution of the initial data (\ref{gauss-exe}) until the catastrophe time $t_c \simeq .78$ estimated from figure \ref{fig-gauss-busur}.}
\label{fig-gauss-evo}
\end{center}
\end{figure}

In figure \ref{fig-gauss-evo} 
 the numerical evolution of the initial data (\ref{fig-gauss-evo}) is shown. We obtain the plots in figure  \ref{fig-gauss-evo}, whose catastrophe 
time is in qualitative agreement with the theoretical prediction.

\subsubsection*{Acknowledgments}
The authors are thankful to E. A. Kuznetsov for attracting their attention to the paper \cite{FKM94}.
G.O. is grateful to  A. Della Vedova  for useful and clarifying discussions. 
This project has received funding from the European Union's Horizon 2020 research and innovation programme under the Marie Sk{\l}odowska-Curie grant no 778010 {\em IPaDEGAN}. We also gratefully acknowledge the auspices of the GNFM Section of INdAM, under which part of this work was carried out, and the financial support of the project MMNLP (Mathematical Methods in Non Linear Physics) of the CSN IV of INFN (Italy). 


\end{document}